\newcounter{myctr}

\documentclass{ws-acs}
\usepackage{url}
\usepackage{graphicx}
\usepackage{amsmath}
\usepackage[german,english]{babel}
\usepackage{epic}
\begin{document}

\makeatletter
\def\@biblabel#1{[#1]}
\makeatother

\markboth{Anders Johansson, Dirk Helbing, Habib Z. Al-Abideen, and Salim Al-Bosta}{From Crowd Dynamics to Crowd Safety}

%
\catchline{}{}{}{}{}
%

\title{From Crowd Dynamics to Crowd Safety: A Video-Based Analysis}

\author{Anders Johansson and Dirk Helbing}

\address{ETH Zurich, UNO D11, Universit{\"a}tstrasse 41, CH-8092 Zurich, Switzerland
\\
andersj@ethz.ch}

\author{Habib Z. Al-Abideen and Salim Al-Bosta}

\address{Central Directorate for Holy Areas Development,\\
Minstry of Municipal and Rural Affairs,\\ Riyadh, Kingdom of Saudi Arabia}

\maketitle

\begin{history}
\end{history}

\begin{abstract}
The study of crowd dynamics is interesting because of the various self-organization phenomena
resulting from the interactions of many pedestrians, which may improve or obstruct their flow. 
Besides formation of lanes of uniform walking direction
and oscillations at bottlenecks at moderate densities, it was recently discovered that stop-and-go waves [D. Helbing {\it et al.}, Phys. Rev. Lett. 97, 168001 (2006)] and a phenomenon called ``crowd turbulence'' 
can occur at high pedestrian densities [D. Helbing {\it et al.}, Phys. Rev. E 75, 046109 (2007)]. 
Although the behavior of pedestrian crowds under extreme conditions is decisive for
the safety of crowds during the access to or egress from mass events as well as
for situations of emergency evacuation, there is still a lack of empirical studies 
of extreme crowding. Therefore, this paper discusses how one may study high-density conditions based on suitable video data. This is illustrated at the example of pilgrim flows entering the previous 
Jamarat Bridge in Mina, 5 kilometers from the Holy Mosque in Makkah, Saudi-Arabia. 
Our results reveal previously unexpected pattern formation phenomena and show that the 
average individual speed does not go to zero even at local densities of 10 persons per square meter. 
Since the maximum density and flow are different from measurements in other countries, this has implications for the capacity assessment and dimensioning of facilities for mass events. When conditions become congested,  the flow drops significantly, which can cause stop-and-go waves and
a further increase of the density until critical crowd conditions are reached.
Then, ``crowd turbulence'' sets in, which may trigger crowd disasters. For this reason,
it is important to operate pedestrian facilities sufficiently below their maximum capacity and to
take measures to improve crowd safety, some of which are discussed in the end.
\end{abstract}


\section{Introduction}

It is well-known that driven many-particle systems often constitute complex systems, in which different kinds of pattern formation phenomena are observed \cite{SomeSuitableBookorReview}. Depending on certain system parameters (the ``order parameters''), one may find transitions from one state of collective behavior to a qualitatively different behavior \cite{HakenSynergetics}. Such transitions typically occur when a certain ``critical threshold'' is crossed. It is interesting to study, whether such transitions can also be found in systems involving humans \cite{Sociodynamics,BrownianAgents}. In this case, the role of particles is replaced by individuals, which follow different interaction rules, Nevertheless, it has been shown that many stylized facts of pedestrian crowds can be well understood by so-called ``social force models'' \cite{TransSci} and other modeling approaches such as cellular automata \cite{Schadschneider,EvacuationWithNagatani}. As examples we mention the segregration of different walking directions into lanes with a uniform direction of motion, or oscillations of the passing direction at bottlenecks \cite{HelbMoln1995}. Recently, we have observed two unexpected transitions in extremely dense pedestrian crowds \cite{PRE_makkah}. While we found laminar flows at small and moderate densities, there was a sudden onset of stop-and-go waves, which was later replaced by a phenomenon of highly irregular motion called ``crowd turbulence''. This dynamics is dangerous, as it may cause people to fall, and it seems to be related with coordination problems between neighboring pedestrians competing for little space. Both, stop-and-go waves and crowd turbulence were previously not expected, because the acceleration time of pedestrians is only about 0.5 seconds \cite{MehdisPNAS}, which suggests a quasi-adiabatic relaxation to a stationary state characterized by the ``fundamental diagram'', i.e. a flow-density relationship. In contrast, stop-and-go waves in freeway traffic are caused by significant delays in the speed adjustments of vehicles \cite{ReviewofModernPhysics}.
\par
The study of dense pedestrian crowds is particularly interesting, since it is one of few systems
involving a large numbers of humans (up to millions), where the detailed dynamics can still be
revealed by analyzing empirical data  \cite{PRE_makkah} or performing agent-based simulations \cite{turbulenceModel}.
Particularly intensive research activities have been triggered by the study of stampedes \cite{panic}.
While ``panic'' has recently been studied in animal 
experiments with mice \cite{mice} and ants \cite{ants} there is still an evident lack of data on critical
conditions in human crowds. Recent work \cite{schadschneider_encyclopedia,rimea} goes into the direction of comparing 
empirical studies with each other, and bringing consensus to some of the fundamental issues in evacuation dynamics. 
\par
However, for a long time, unidirectional pedestrian flows were predominantly assumed to move 
smoothly according to the "fluid-dynamic" flow-density relationship \cite{Weidmann}
\begin{equation}
Q(\rho) = \rho V(\rho)\, ,
\label{fluidform}
\end{equation} 
where $Q$ represents the flow per meter width, $\rho$ is the pedestrian density, and the average velocity $V$ 
is believed to go to zero at some maximum density as in traffic jams \cite{Fruin2,Predtechenskii,Mori,Polus,Weidmann,Seyfried}.
Formula (\ref{fluidform}) is often used as a basis for dimensioning and designing pedestrian facilities, 
for safety and evacuation studies. However, empirical measurements are often restricted to densities
up to 4--6 persons per square meter only. In the study \cite{Weidmann}, for 
example, the maximum density $\rho_{\rm max}$ is 5.4 persons per square meter, and 
the corresponding fit curve of the speed-density relationship is
\begin{equation}
 V(\rho) = V_0
 \left\{ 1 - \exp \left[ -a \left(\frac{1}{\rho} - \frac{1}{\rho_{\rm max}} \right)\right]\right\} \, ,
 \label{WEID}
\end{equation}
where $V_0 = 1.34$m/s is the free speed at low densities and $a = 1.913$ persons per square meter
a fit parameter. Some other measurements of pedestrian densities, 
however, reach upto 6 persons per square meter or more \cite{Predtechenskii,Mori}. So, which measurement is correct?
And what happens at even higher densities? Based on the projected area of human bodies,
upto 11 persons may fit into one square meter \cite{Still}.
\par
In order to address these issues, we will evaluate video recordings of the annual Muslim pilgrimage,
where conditions are known to be particularly crowded. We will also see that these data are not 
directly transferable to Western European conditions and vice versa. 
Video-based studies of pedestrian flows have recently been carried out by other authors
as well \cite{Kretz,hoogendoorn2005,Seyfried}. Most
of them focus on bottleneck rather than unconstrained flows. Helbing {\em et al.} \cite{BottleneckPRL} have recently
proposed a macroscopic model for such flows. The situation we analyze in the following, however,
is characterized by particularly extreme densities, which have finally led to the crowd disaster
on January 12, 2006. Therefore, we expect particular insights into critical crowd conditions.
\par
Our paper intends to identify some reasons for 
the different flow-density curves (fundamental diagrams) reported in the
literature and to close the data gap in the safety-relevant range of extreme densities.
Considering the inconsistent measurements in the literature, which one
should a capacity analysis be based on? Using
the wrong curves can easily imply a wrong dimensioning and, thereby,
future disasters. Generally, the difference between local and global measurements has
not been sufficiently paid attention to. The same applies to the body size
distribution and the cultural backgrounds (regardings prefered spacing and
speeds). Many papers also don't present standard deviations, which are
important to judge the variability of the measurement points, to compare
different data sets and to determine the required capacity reserves when
planning pedestrian facilities. 
\par
The technique of video analysis is an important, but not the main
point of our paper. Nevertheless, we show in detail how certain
problems related to video-based evaluations can be successfully overcome
and how the results depend on the specification of parameters. Note that the video-based
evaluation method described in this paper was also used for another study
\cite{PRE_makkah}. While that study focussed on the {\em dynamics} of
crowd disasters, this paper addresses the {\em measurement process} and 
{\em safety-relevant features} of the speed-density and flow-density diagrams.
These issues are important to draw correct conclusions
from the video data. Another focus are the determination of critical crowd 
conditions. It will turn out that neither the density nor the speed or flow
field are good measures of the criticality in the crowd, as the latter actually depends on the
crowd {\it dynamics}.
\par
The remainder of this paper descibes the large efforts and various difficulties that must be overcome in order to
extract the characteristics and dynamics of crowd behavior at large densities from video recordings.
As commercial tracking software did not do the job, we had to develop new algorithms
which were capable of dealing with hundreds rather than dozens of people in a fully
automated way. This involved the evaluation of terabytes of data, recorded  at many measurement sites over several days
in January 2006 (during the Hajj of the year 1426H). The whole project,
including the comparison with manual counts for validation and calibration of the method
required several man years.
\par
Our paper is structured as follows: Section~\ref{video} will describe our technique of
video evaluation, which is the basis of our data analysis. In Sec.~\ref{measure} we 
will then define a new measurement method for local densities and flows. Problems
like how to correct for the hiding of people by umbrellas will be addressed as
well. Afterwards, Sec.~\ref{fundamental} will present empirical 
measurement results for the flow-density and speed density diagrams
at the entrance of the previous Jamarat Bridge. Here, we will also study the 
considerable variability of density, velocity and flow data. 
In Sec.~\ref{critic} we will describe the stop-and-go waves and the phenomenon of  ``crowd turbulence''
discovered at high densities. Moreover, we will
determine warning signs of critical crowd conditions, in particular the ``crowd pressure''.
Section~\ref{summary} will summarize our results and discuss
implications for the safety analysis and dimensioning of pedestrian facilities.

\section{Measurement Site and  Video Tracking}
\label{video}

\subsection{Description of Measurement Site and Conditions}

\begin{figure}[!htbp]
\begin{center}
\includegraphics[width=1.0\textwidth]{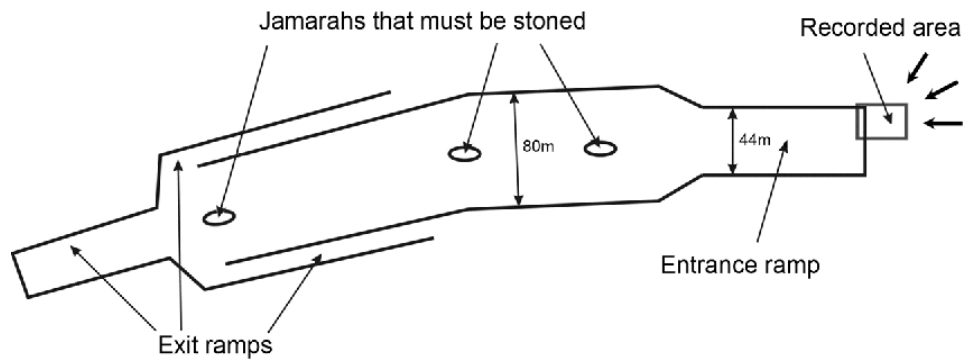}\,
\caption[]{
Illustration of the old Jamarat Bridge and the video-recorded area we are concentrating on
in this study. There were many more cameras installed, 
but the one in the entrance area showed the highest
densities and the most interesting crowd dynamics.}
\label{Fig0}
\end{center}
\end{figure}

In order to get a better understanding of extremely dense conditions, we have scientifically evaluated
the situation during the stoning ritual in Mina through 12 fixed cameras mounted on high poles.
The overall video material of the 10th to 12th day of Dhu al-Hijjah, 1426H amounts to more than 2 Terabyte of data. 
A special focus of the study presented here will be on 
the entrance area of the Jamarat Bridge (see Fig.~\ref{Fig0}), where the situation became
particularly crowded and the sad crowd disaster
happened on January 12, 2006 (the 12th day of the pilgrimage). 
\par

\subsection{Video Tracking Method}

Video tracking has become a common and comfortable tool for empirical pedestrian
research recently \cite{Teknomo,HoogendoornEmpirical,Kerridge}.
Its big advantage is the automatic evaluation of pedestrian trajectories,
but video tracking has also its limitations:
\begin{itemize}
\item Some tracking softwares require to specify the starting points of pedestrians
manually by clicking on their heads.
\item Possible camera positions often imply small areas of recording or
oblique camera positions. The latter implies that pedestrians may be hidden behind
each other at high densities. In both cases, one must correct for effects of perspective.
\item In dense pedestrian crowds, pedestrians are often ``lost'' or interchanged by
the tracking algorithm.
\item Most tracking algorithms are restricted to tracking several dozens of pedestrians
for the above reasons and reasons of numerical performance.
\end{itemize}
Thanks to the 35 meter high pole at which our video camera was mounted, the recorded
area was quite large (28m$\times$23m) and
the recordings could be made almost perpendicular to the ground. Therefore, pedestrians 
of different height were usually not hidden behind each other. However, the radius of a pedestrian
head extended over 2--3 pixels only (corresponding to about 15--25 pixels covered by each head).
\par\begin{figure}[!htbp]
\begin{center}
\includegraphics[width=0.5\textwidth]{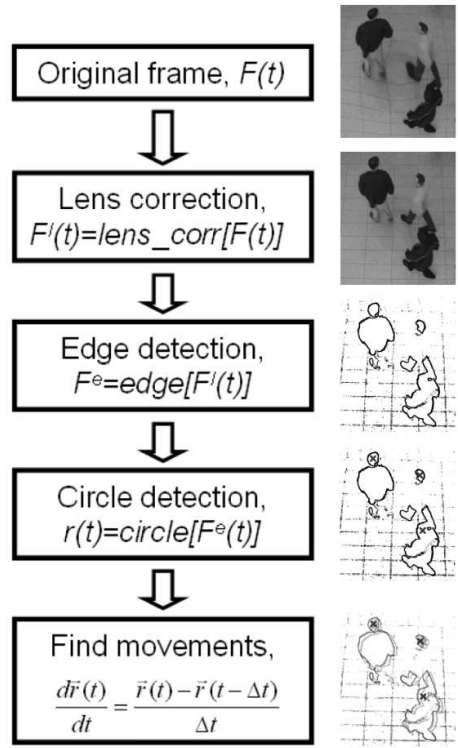}
\end{center}
\caption[]{Illustration of how the video-analysis software operates. The raw video data are fed from 
the top, and each frame is passing a number of filters (actually more than shown over here), until the head locations and velocities are obtained with sufficient reliability in the end.} 
\label{video_tracking_loop}
\end{figure}
\begin{figure}[!htbp]
\begin{center}
\includegraphics[width=1.0\textwidth]{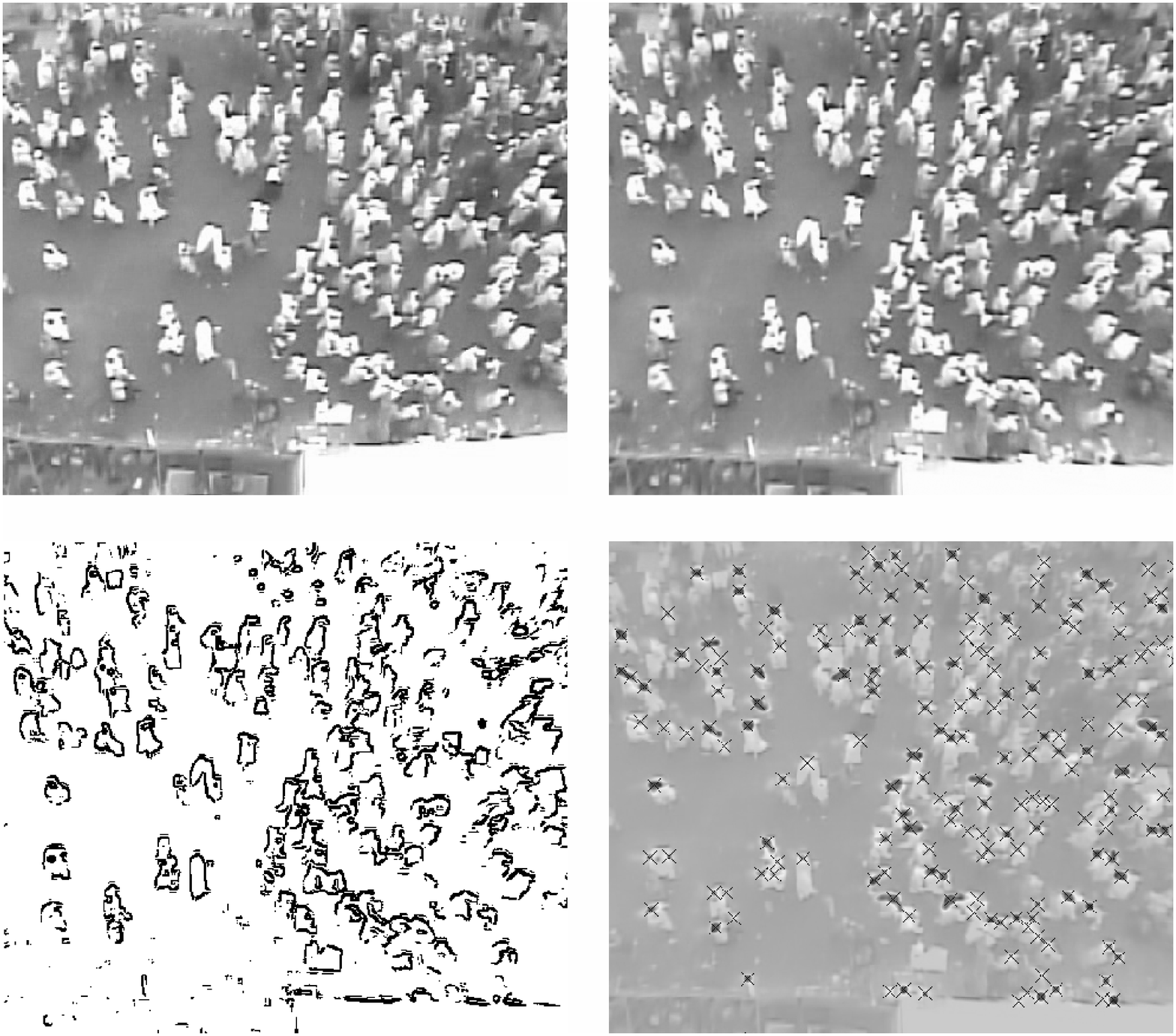}
\end{center}
\caption[]{Illustration of video frames before and after certain transformations performed by our video analysis software. Upper left: Original frame. Upper right: Same frame, after the lense distortion was corrected for (see the straightened wall in the lower part.) 
Bottom left: The frame after edge detection and thresholding were applied.
Bottom right: Frame superposed to the head-detection probability-density surface, with crosses on the most likely head locations as determined by an Artificial Neural Network (ANN) \cite{johansson_phd}.
The probability density is determined by searching for head-like patterns in the current frame as well as extrapolating the locations of head detections made in previous frames, which results in a reinforcement when the same head is detected in many consequitive frames.
A cross is displayed if the probability density is above a certain threshold and there is no more likely head location within a distance corresponding to one head diameter. The inaccuracies in the upper left corner are both due to the fact that the video is blurry in this region, but also because these pedestrians have just entered the video and therefore the reinforcement of
the tracking algorithm has not yet taken effect.}
\label{tracking_frames}
\end{figure}
\begin{figure}[!htbp]
\begin{center}
\includegraphics[width=1.0\textwidth]{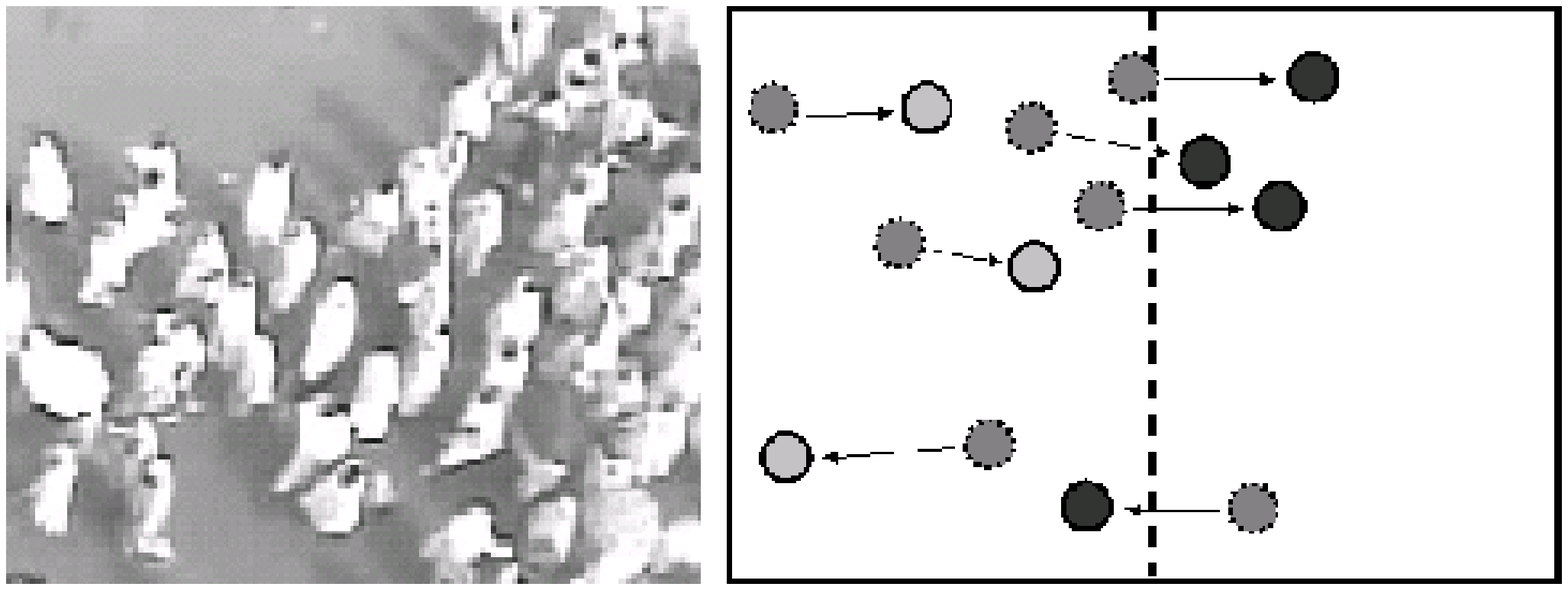}
\end{center}
\caption[]{Left: Part of a video frame used for manual counting and calibration.
Right: Schematic illustration of the method used for the manual counting of pedestrians.
The crowd video is played for 5 seconds in slow-motion at 1/10th of the original speed, and two people are independently counting
the number of pedestrians who cross a certain line within these 5 seconds. The counting is not done for the complete video recording, but for short video sequences separated by time intervals of 10 minutes each.}
\label{heads}
\end{figure}

The software developed by ourselves automatically determines heads by seaching for round structures.
For this purpose, several digital filters (transformations) are successively applied to the video frames
to enhance their contrast and identify the relevant structures (see Fig. \ref{video_tracking_loop}). 
To gain a higher accuracy, the following additional steps are performed: 
\begin{itemize}
\item Double-checking of the identified circles by applying an Artificial Neural Network \cite{johansson_phd} trained to recognize heads.
\item Application of adaptive histogram equalization to compensate for variations in the light conditions.
\item Application of corrections derived from extensive manual count data (see Fig. \ref{heads} and Sec.~\ref{manual_counts}).
\end{itemize}
Figure~\ref{tracking_frames} illustrates how the 
video frames look like during several processing steps of the tracking algorithm.
\par

For each frame, the identification of heads yields the locations $\vec{r}_i(t)$ of the pedestrians $i$ at time $t$.
By comparing these with their local neighborhoods in the next video frame, one can determine their velocities $\vec{v}_i(t)$.
The speed information is used to estimate the location of the pedestrians in the subsequent frame,
which improves the accuracy of the tracking procedure.
With a resolution of 25 pixels per meter and
8 frames per second, it is possible to determine even small average speeds by
calculating the mean value over a large enough sample of individual speed measurements.

\par

\subsubsection{Comparison of Automated and Manual Counts}
\label{manual_counts}

\begin{figure}[!htbp]
\begin{center}
\includegraphics[width=0.32\textwidth,angle=-90]{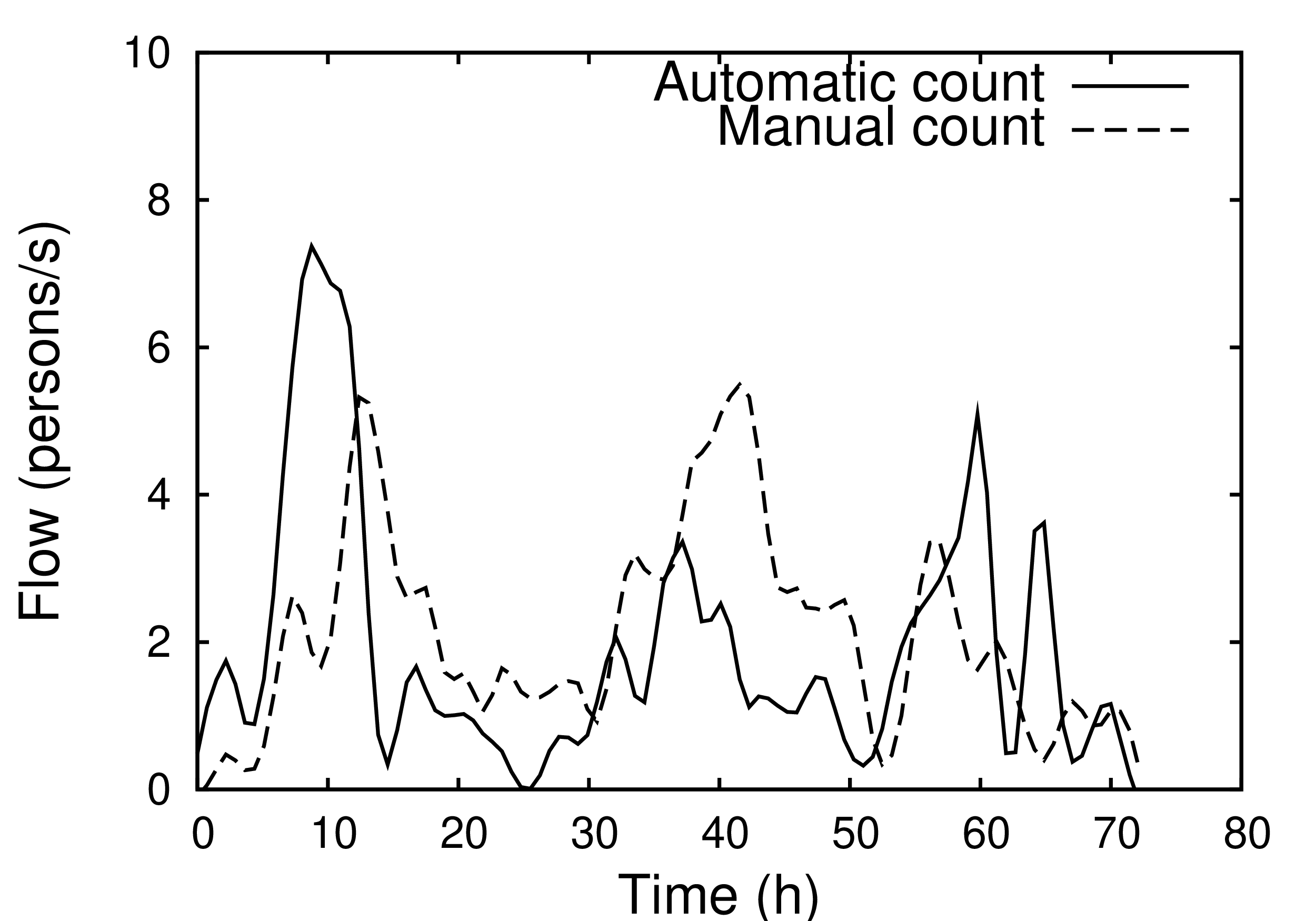}
\includegraphics[width=0.32\textwidth,angle=-90]{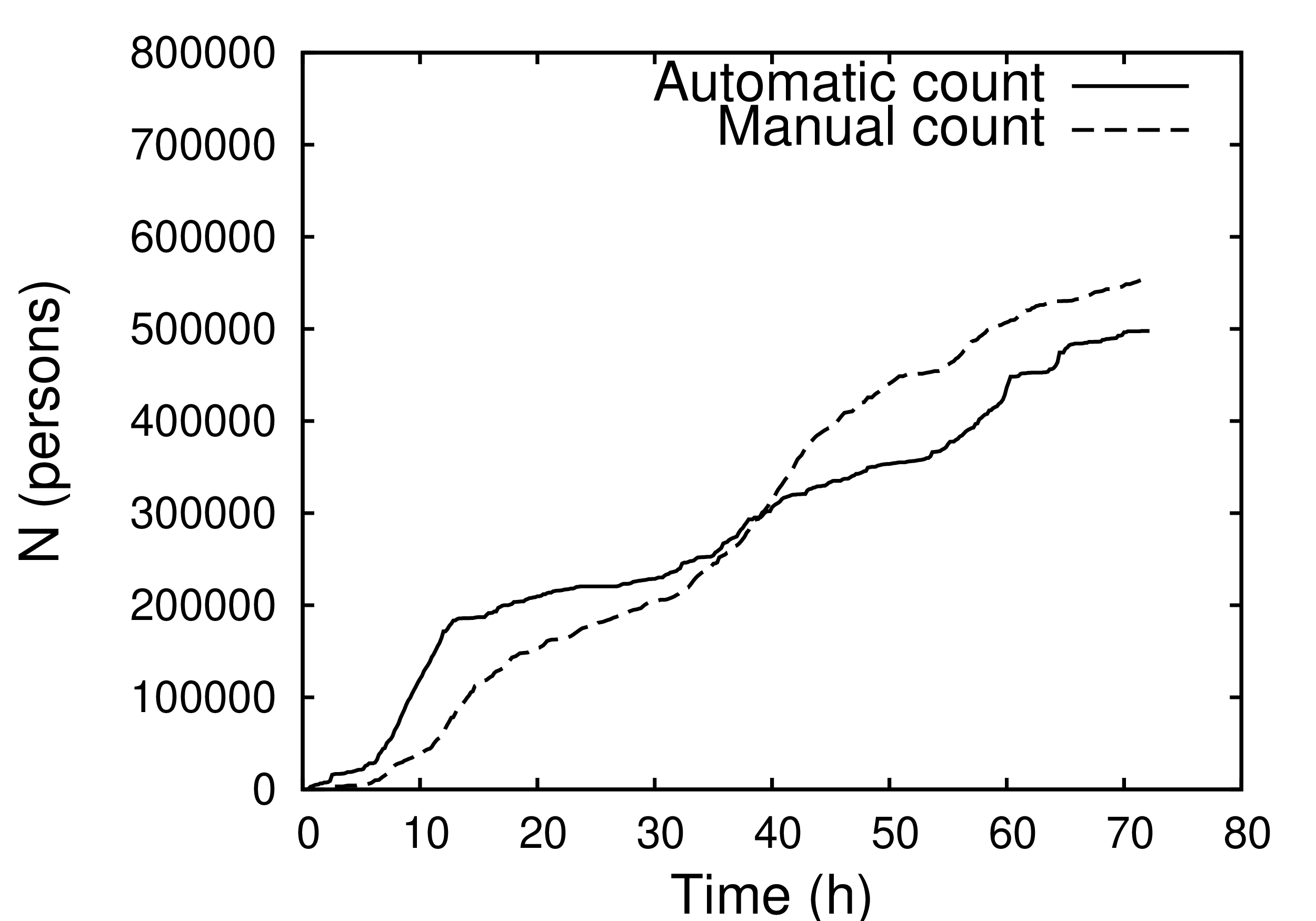}\\
\includegraphics[width=0.32\textwidth,angle=-90]{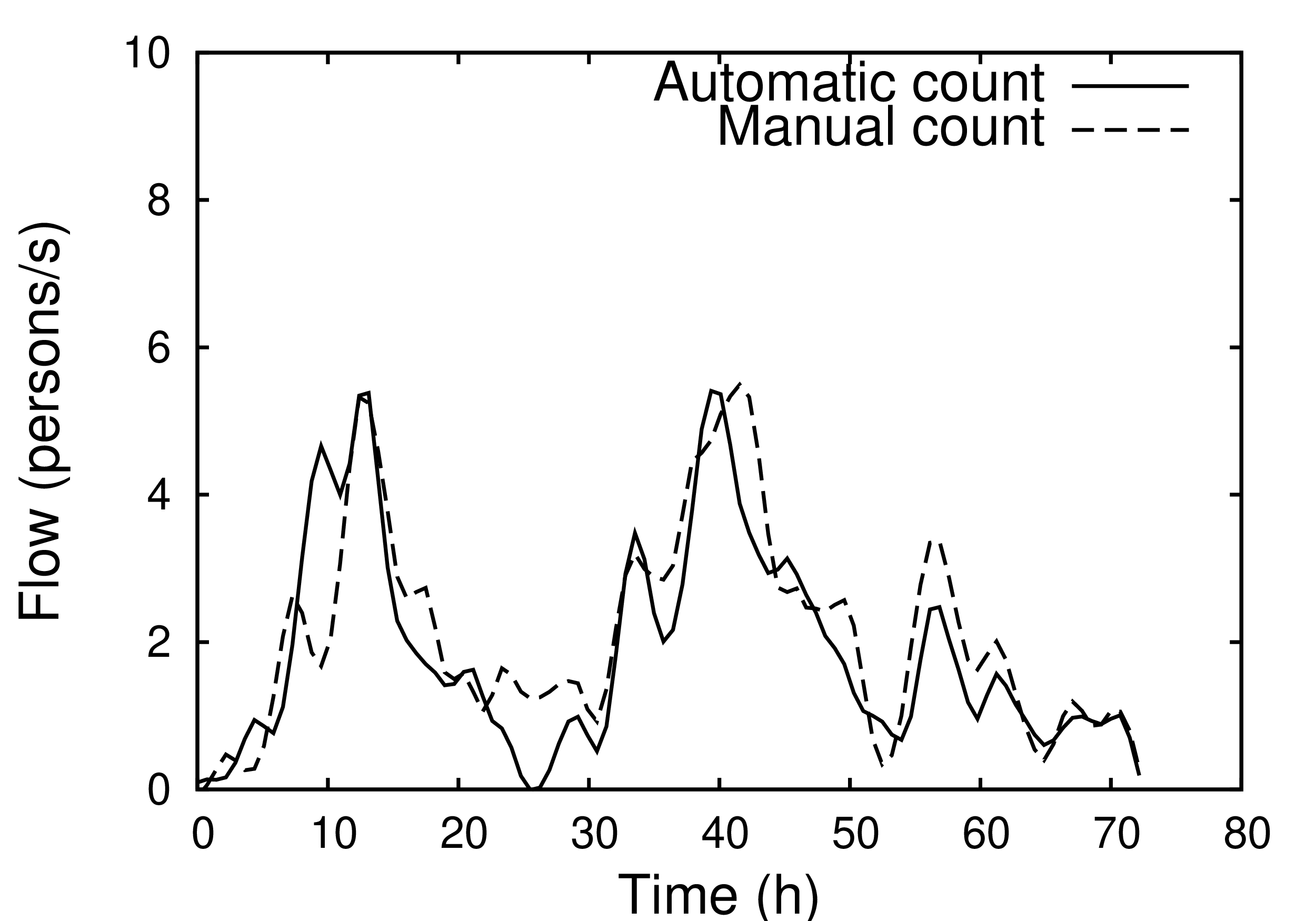}
\includegraphics[width=0.32\textwidth,angle=-90]{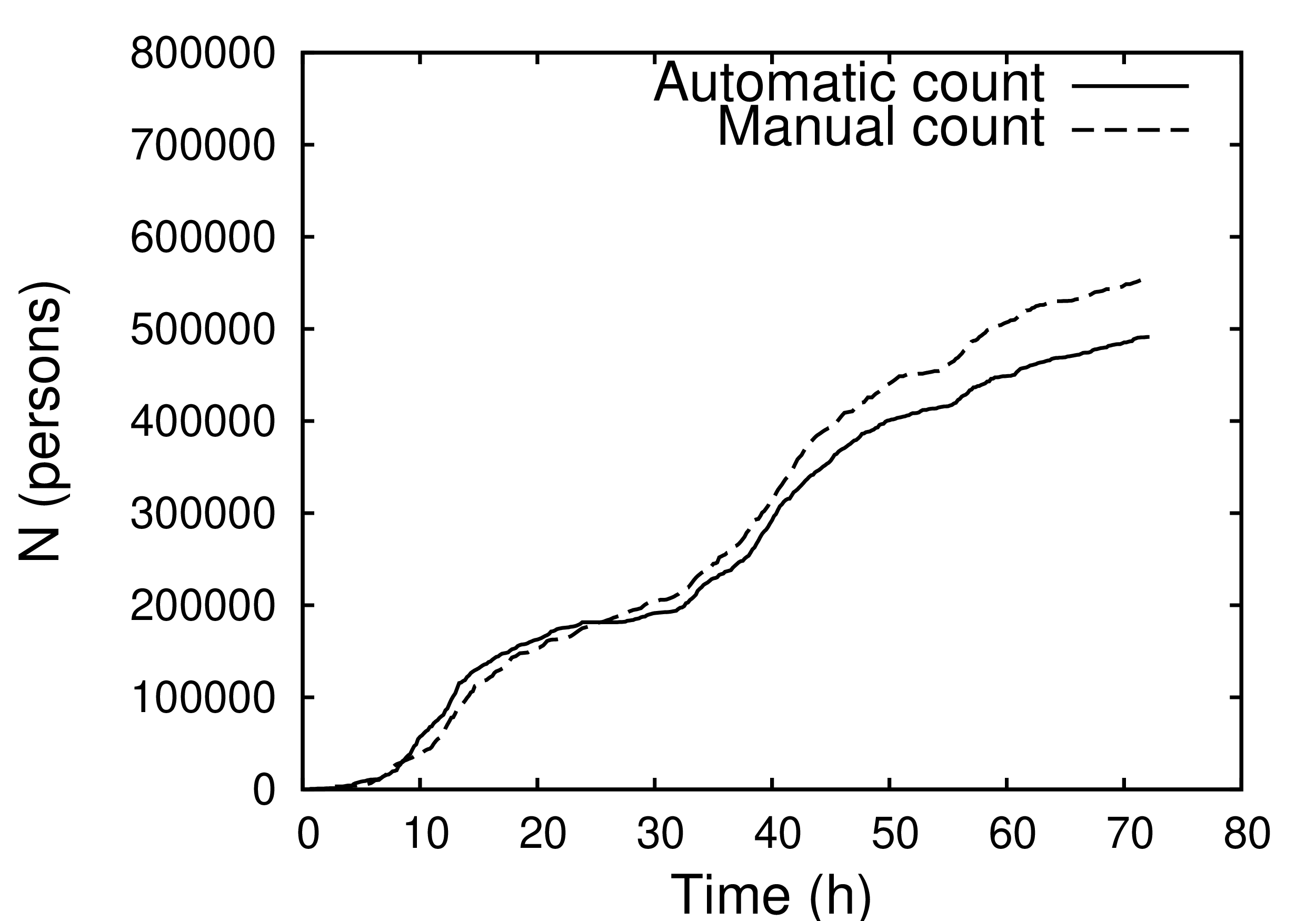}
\end{center}
\caption[]{Flow and cumulative flow of a pedestrian road
to the Jamarat plaza over 72 hours. Top: Comparison of automatically and
manually determined values, using the methods illustrated in Figs.~\ref{video_tracking_loop} and \ref{heads}.
Bottom: Same comparison, when more sophisticated classifiers based on Artificial Neural Networks \cite{johansson_phd}
were used.}
\label{comparison}
\end{figure}

The tracking routine does not only give good estimates of the local densities, speeds and flows.
It is also suitable for counting people. Figure~\ref{comparison} compares 
automated and manual counts for one street leading to the Jamarat Plaza. Despite the difficulty to
distinguish people walking in opposite directions, the reliability of the automated counts 
is reasonably high. The deviation is actually of the same order as the deviation between 
manual counts of two different persons.
\par

\section{Measurement of Local Densities, Speeds, and Flows}
\label{measure}

\subsection{Data Evaluation Method}

From the locations $\vec{r}_i(t)$ of the pedestrians $i$ at time $t$, we
have determined the local density $\rho(\vec{r},t) = 
\rho_t^R(\vec{r})$ at a location $\vec{r}$ via the formula
\begin{equation}
\rho(\vec{r},t) = \rho_t^R(\vec{r}) 
= \frac{1}{\pi R^2} \sum_j \exp[-\|\vec{r}_j(t) - \vec{r}\|^2/R^2] \, ,
\label{dens}
\end{equation} 
where $R$ is a parameter. 
The {\it local} velocities have been defined via the weighted average 
\begin{equation}
 \vec{v}(\vec{r},t)  = \vec{v}_t^R(\vec{r}) = \frac{\sum_j \vec{v}_j \exp[-\|\vec{r}_j(t) - \vec{r}\|^2/R^2]} 
{\sum_j \exp[-\|\vec{r}_j(t) - \vec{r}\|^2/R^2] } 
\end{equation}
and {\it local} speeds as $v(\vec{r},t) = v_t(\vec{r}) = \|\vec{v}_t^R(\vec{r})\|$, 
while {\it local} flows have been obtained according to the fluid-dynamic formula
\begin{equation}
 q(\vec{r},t) = q_t^R(\vec{r}) = \rho(\vec{r},t)\vec{v}(\vec{r},t) \, . 
\end{equation}
The greater $R$, the greater the smoothing area around $\vec{r}$. 
It can be calculated that the weight of neighboring pedestrians located within
the area $A_R = \pi R^2$ of radius $R$ is 63\%. 
\par
In principle, the average {\em (``global'')} values $\varrho(t)$, $V(t)$, and $Q(t)$
of the density, speed, and flow, respectively, can be determined via the formulas for $\rho_t^R(\vec{r})$, $v_t^R(\vec{r})$ 
and $q_t^R(\vec{r})$, if the value of $R$ is chosen sufficiently large. However, $R$ 
must be significantly smaller than the radius of the video-recorded area in order to avoid 
boundary effects. Moreover, $\vec{r}$ must be chosen from the central part of the video so that the areas $A_R$ 
around $\vec{r}$ are completely recorded. In practise, the global density is measured by counting all pedestrians in the 
area of interest and then dividing by that area. The global velocity is determined by 
calculating the average velocity of 
all persons in that area, and the global flow is defined by the global density 
times the global velocity, or by the number of people passing a long cross section per unit time,
divided by its length.
\par
\begin{figure}[!htbp]
\begin{center}
\includegraphics[width=1.0\textwidth]{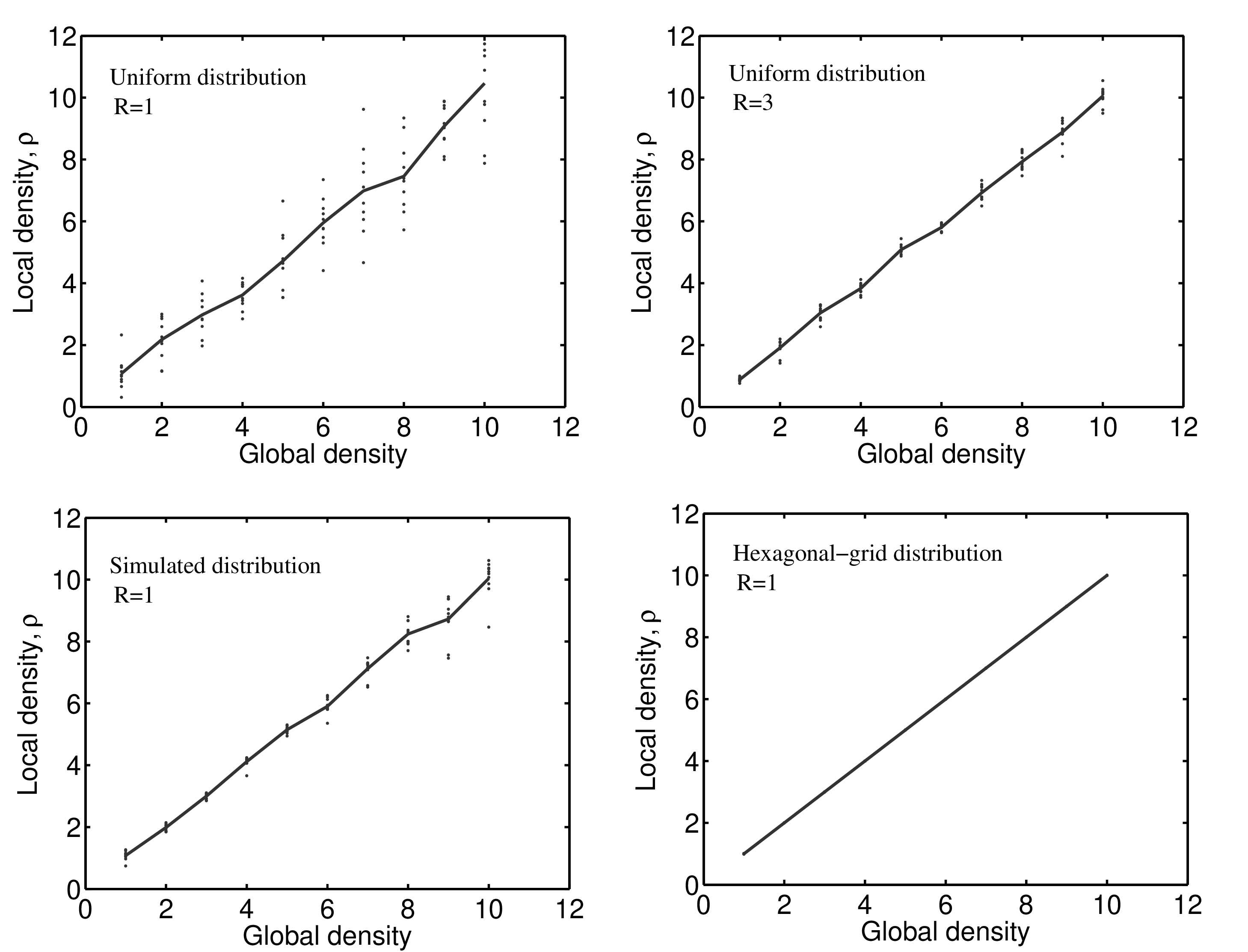}
\caption[]{Local density measurements according to formula (\ref{dens}) for 10 randomly
picked points in a circle of 10 meter radius, compared
to the average density in a circular area $A_R$ of radius $R=100$ meters.
Top: Uniformly distributed points with $R = 1$ (left) and $R = 3$ (right). Note that the local density
of randomly distributed points varies strongly, as different points can be arbitrarily close to each other. 
Therefore, we also generated pedestrian distributions resulting from 
a pedestrian simulation with the social force model \cite{TransSci}, which took into account the finite space requirements (``diameters'') of pedestrians. In this simulation, we assumed an average desired velocity of 1.34 meters per second and a standard deviation of 0.26 meters per second as in Ref.~\cite{Weidmann}. The desired direction of walking were assumed to be the same for all pedestrians. The initial distribution was randomly chosen, but the density measurement was
made after a simulated time of 30 seconds. The resulting density distribution for $R=1$ (bottom left) is significantly smaller, as the distribution of pedestrians is more regular than a random distribution due to their repulsive interactions. 
To a certain extent, it reminds of the distribution of points located on a hexagonal lattice 
(bottom right, for $R=1$).}
\label{Fig1}
\end{center}
\end{figure}
In Fig.~\ref{Fig1} we have studied
by means of computer-generated distributions of pedestrians, 
how well the average (``global'') density values $\varrho$ are reproduced for different values 
of $R$. Our evaluation gives the following: The average $\langle \rho_t^R\rangle$  of the local density values $\rho_t^R$ 
according to Eq. (\ref{dens}) over different locations $\vec{r}$ 
agrees well with the average (``global'') density $\varrho$. Moreover, the variance of the
local density measurements around the given, average density $\varrho$, 
goes down with larger values of $R$. 
In fact, for $R \rightarrow \infty$, all local density measurements result in the same value
\begin{equation}
\varrho = \lim_{R\rightarrow \infty} \rho_t^R(\vec{r}_i) \, .
\end{equation}
It corresponds exactly to the overall number $N_R$ of pedestrians, divided by the area $A_R = \pi R^2$ they are distributed in, i.e.
\begin{equation}
 \lim_{R\rightarrow \infty} \left( \rho_t^R(\vec{r}_i) - \frac{N_R}{A_R} \right) = 0 \, .
\end{equation}
Division of $N_R$ by $A_R$ corresponds to the classical method of determining the
average density $\varrho$, but it can also be obtained by averaging over local density measurements $\rho_t^R(\vec{r})$:
\begin{equation}
 \frac{1}{\pi R^2} \sum_j \int \exp[-\|\vec{r}_j(t) - \vec{r}\|^2/R^2] \, d^2r  
=  \frac{1}{A_R} \sum_j  \int\limits_0^{2\pi} \!\int\limits_0^\infty \!   
 \mbox{e}^{-r^2/R^2} r\, d\varphi \, dr 
= \frac{\sum_j 1}{A_R} = \frac{N_R}{A_R} \, . 
\end{equation}
Note, however, that we are not so much interested here in the average density $\varrho$, but in the local 
density $\rho(\vec{r}_i,t)$, as this is expected to determine the behaviour of pedestrian $i$. The variation of the values of $\rho_i(\vec{r},t)$ in Fig.~\ref{Fig1} is due to the statistical variation 
of pedestrians in space, which results in local density variations. If the value of $R$ is fixed,
the relative variation of measured density values is smaller for higher densities, due to the larger number
of pedestrians in the area $N_R = \pi R^2$. This is a favorable property of our local density measurement
method, as we are particularly interested in high densities.
\par
When measuring
the local densities from video recordings of crowds, there may be additional reasons for
density variations:
\begin{enumerate}
\item errors in the automated identification of pedestrians due to bad video quality
(or, in cases of oblique cameras, due to hiding of other persons),
\item hiding of persons below umbrellas (to protect themselves against the sun), and
\item variations due to the spatio-temporal dynamics of the crowd.
\end{enumerate}
While the first problem can be partially healed by tracking over certain time periods,
the second point will be addressed next. The third point was the subject of another
publication \cite{PRE_makkah}.

\subsection{Dealing with Umbrellas} \label{sec_umbrellas}

Before we proceed with our evaluation, we have to discuss one particularity of our video recordings: 
A certain fraction $p$ of pilgrims uses an umbrella for the protection from sun. These umbrellas mostly
have a radius of 50 cm, so they can be used for calibration. However, they also cover an unknown
number of pilgrims. In the determination of the average density $\varrho$, one can correct 
for umbrellas as follows: Let $A$ be the area covered by the video recording, 
$A_1=A/N$ the area available for each of the $N$ persons in this area, and $A_2$ the typical area of an umbrella. 
Then, one umbrella corresponds to $k = A_2/A_1$ covered persons.
Moreover, let $n$ be the number of
visible people in the area not covered by umbrellas and $m$ the number
of umbrellas. The overall number of people is then $N = n + km$. If we
assume that each $p$-th pedestrian carries an umbrella, we have $m = pN$,
i.e. $N = n + kpN$ or $N = n / (1 - kp) = n / (1 - pA_2/A_1)$.
The correct average density is $\varrho = N / A$, while the density calculated without
considering umbrellas is $\varrho' = n / A$. Therefore,
\begin{equation}
\varrho = \frac{\varrho'}{1 - pA_2/A_1} = \frac{\varrho'}{1 - p\varrho A_2} \, ,
\label{eq_p}
\end{equation}
as $\varrho = 1 / A_1 = N/A$. That is we would not have any corrections for $p=0$, $\varrho=0$, or
$A_2=0$. However, in reality corrections are needed. These lead, on average, 
to higher density values. The corrected value for $\varrho$ can be calculated as solution of a quadratic
equation. We get
\begin{equation}
\varrho = \frac{1 - \sqrt{1 - 4pA_2\varrho'}}{2 p A_2} \, . 
\label{VARRHO}
\end{equation}
While this solves the problem of how to correct the {\it average} density, if the fraction $p$ and average area $A_2$ of
umbrellas is known, it does not help to correct the {\it local} density measurements, as the fraction of 
the measurement area covered by umbrellas is varying {\it strongly}. This can lead to 
significant variations in the local density measurements, even if the local density is actually the same.
\par
\begin{figure}[!htbp]
\begin{center}
\includegraphics[width=0.49\textwidth]{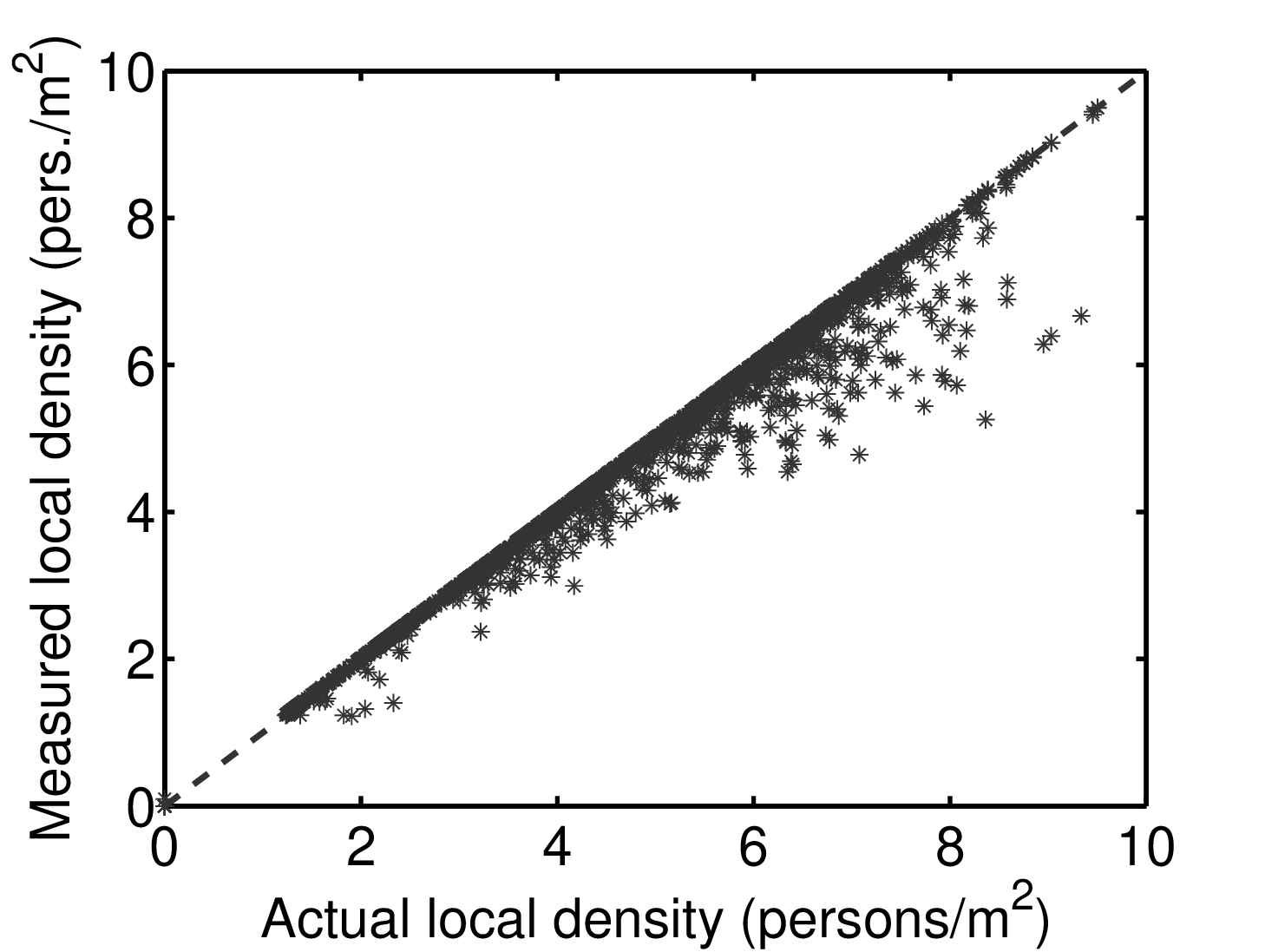}\, 
\includegraphics[width=0.49\textwidth]{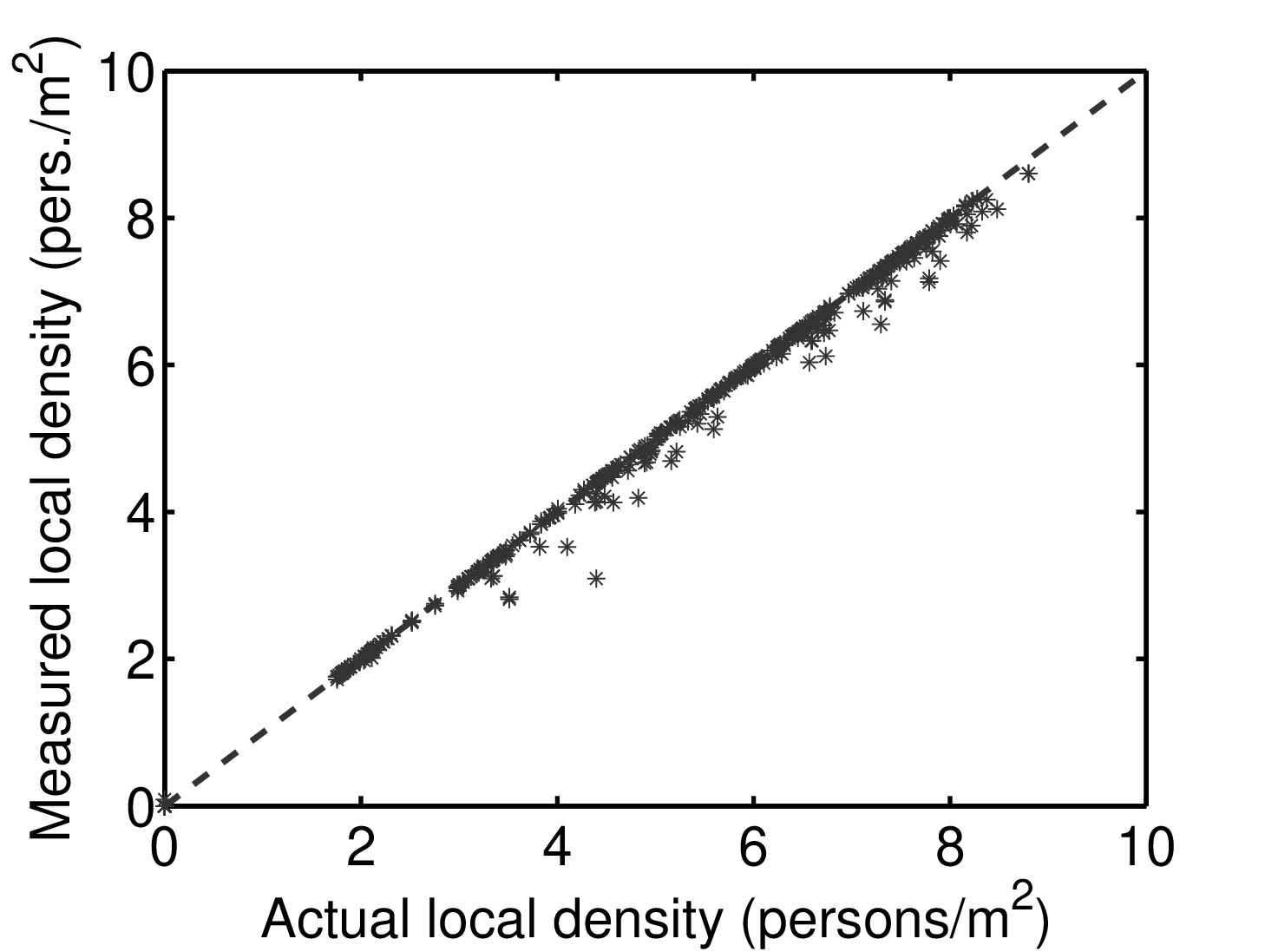}
\caption[]{Simulated results to determine the influence of umbrellas on the local density measurements.
The computer simulations have been carried out with the social force model, starting with random initial conditions. 
We have assumed that 2\% of the persons carry umbrellas with a radius of 50 cm.
While the {\it actual} local density has been determined from all generated points (representing pedestrians),
the {\it measured} local density has been determined by removing all points in a radius of 50 cm around 
2\% of the points. The figures show the measured over the actual local density for a cutoff of
$\gamma = 0.5$ (left) and $\gamma = 0.95$ (right). A cutoff value of $\gamma = 0.95$ reproduces local density measurements over a large range of densities well, while a cutoff of $\gamma = 0.5$ tends to underestimate the actual local densities.}\label{umbrellafig}
\end{center}
\end{figure}

For that reason, within a 20m$\times$15m area, we have randomly picked pedestrian locations
and have determined the local densities around them. From this sample, we have removed
the fraction $\gamma$ of lowest density values. 
Since the umbrellas are most likely to be found in the low-density regime, the probability
of successfully filtering out umbrellas grows with $\gamma$. 
Note that we do not assume that persons with umbrellas tend to
avoid large densites. Rather the idea is that the density calculation 
under-estimates the density in the vicinity of
an umbrella because the surrounding persons are hidden under the umbrella. 
Fig.~\ref{umbrellafig} shows for computer-generated data that filtering out 50\% of
the low-density data can still lead to significant under-estimation of the actual density 
by the measured one, while $\gamma = 0.95$ leads to reasonably accurate density measurements.
Therefore, in the following empirical evaluations of video-recorded pilgrim flows,
we will restrict to sufficiently reliable measurements of local densities 
by using a cutoff value of $\gamma = 0.95$.
\par
\begin{figure}[!htbp]
\begin{center}
\includegraphics[width=0.49\textwidth]{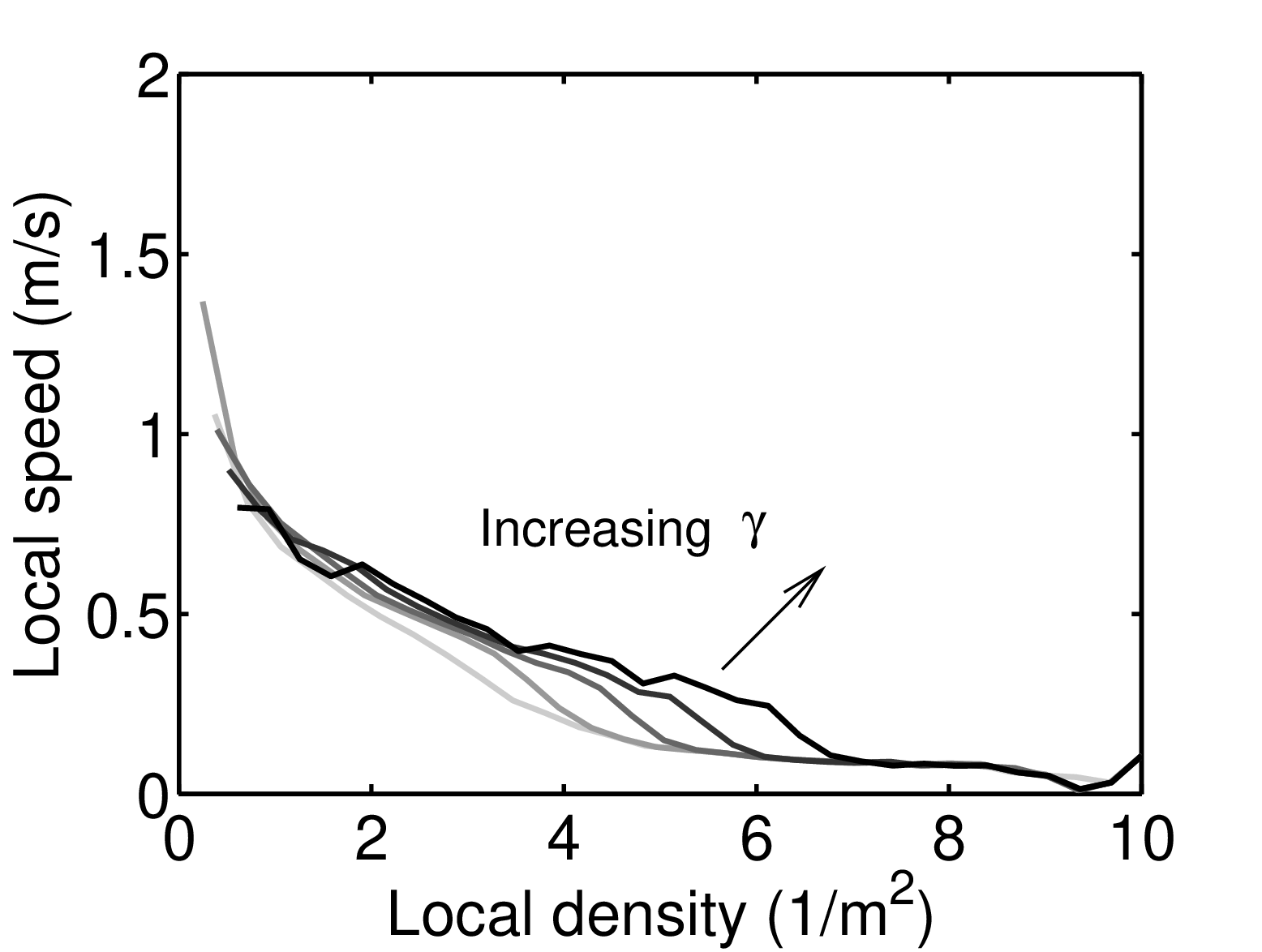}\,
\includegraphics[width=0.49\textwidth]{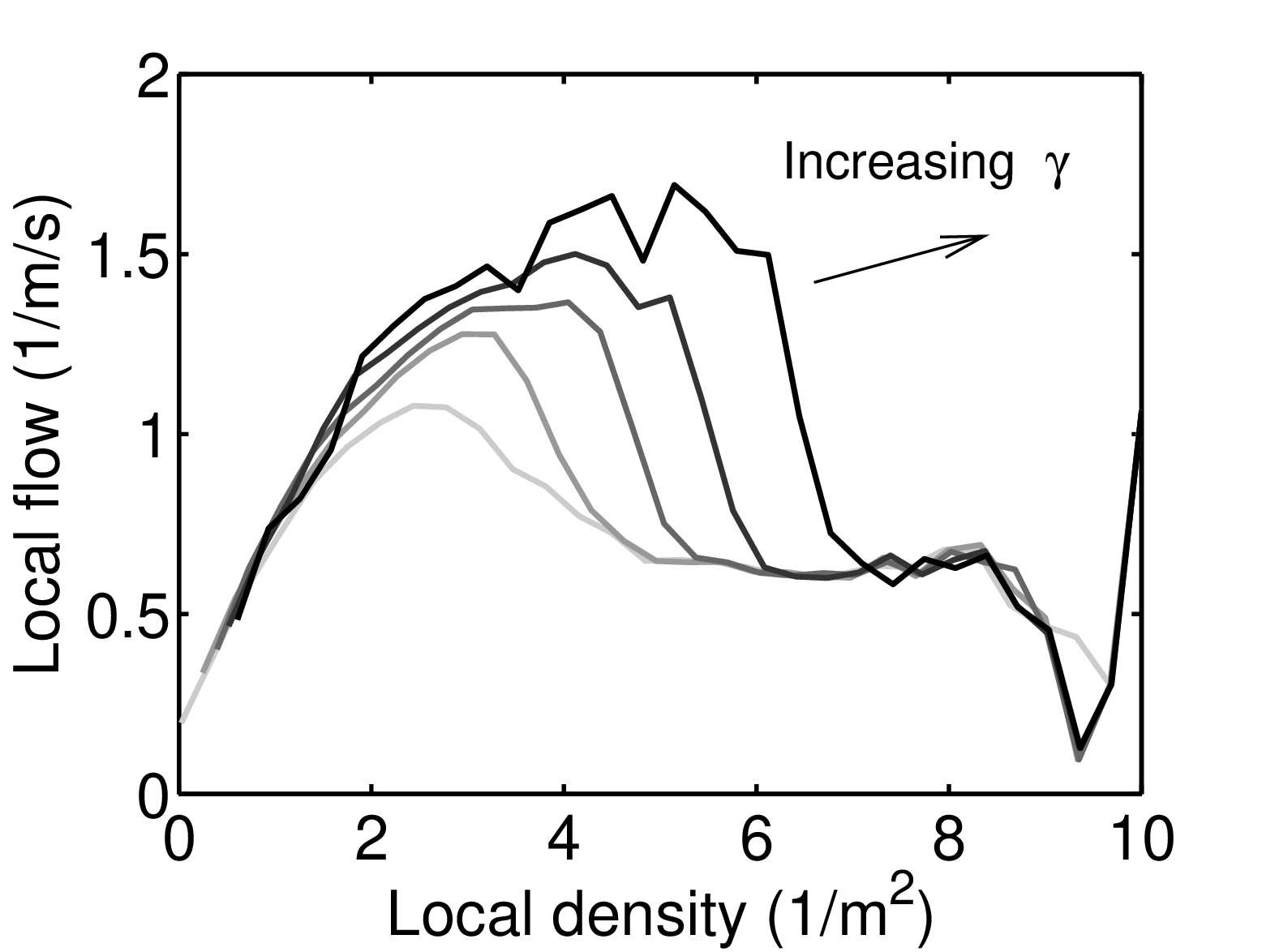}\,
\caption[]{Empirical relationships 
between the local speed and the local density (left) and 
between the local flow and the local density (right) 
for the measurement site depicted in Fig.~\ref{Fig0} and various values of $\gamma \in \{
0\%, 25\%, 50\%, 75\%, 95\%\}$. The cutoff value $\gamma$ is supposed to 
correct for the influence of umbrellas. It is clear that the flow values are underestimated if pedestrians
covered by umbrellas are not suitably accounted for, as for small values of $\gamma$.}
\label{gamma_dependence}
\end{center}
\end{figure}

Let us now check the plausibility of the above proposed methods to correct for umbrellas with real
data. In order to investigate how the speed-density and flow-density diagrams depend on the cutoff value 
$\gamma$, we have evaluated them for different values of $\gamma$
(see Fig. \ref{gamma_dependence}). We find that, while the speed-density relationships vary relatively little
with the variation of $\gamma$, there is a significant increase of the local flows for greater $\gamma$ values,
particularly in the intermediate density range. This shows that fitting speed-density data may lead to
unreliable conclusions regarding the flow-density relationship, while fitting flow-density data would
lead to good velocity-density fits as well. The reason is that the density enters twice and in a multiplicative
manner into the flow (as product of density and speed).
\par
\begin{figure}[!htbp]
\begin{center}
\includegraphics[width=0.49\textwidth]{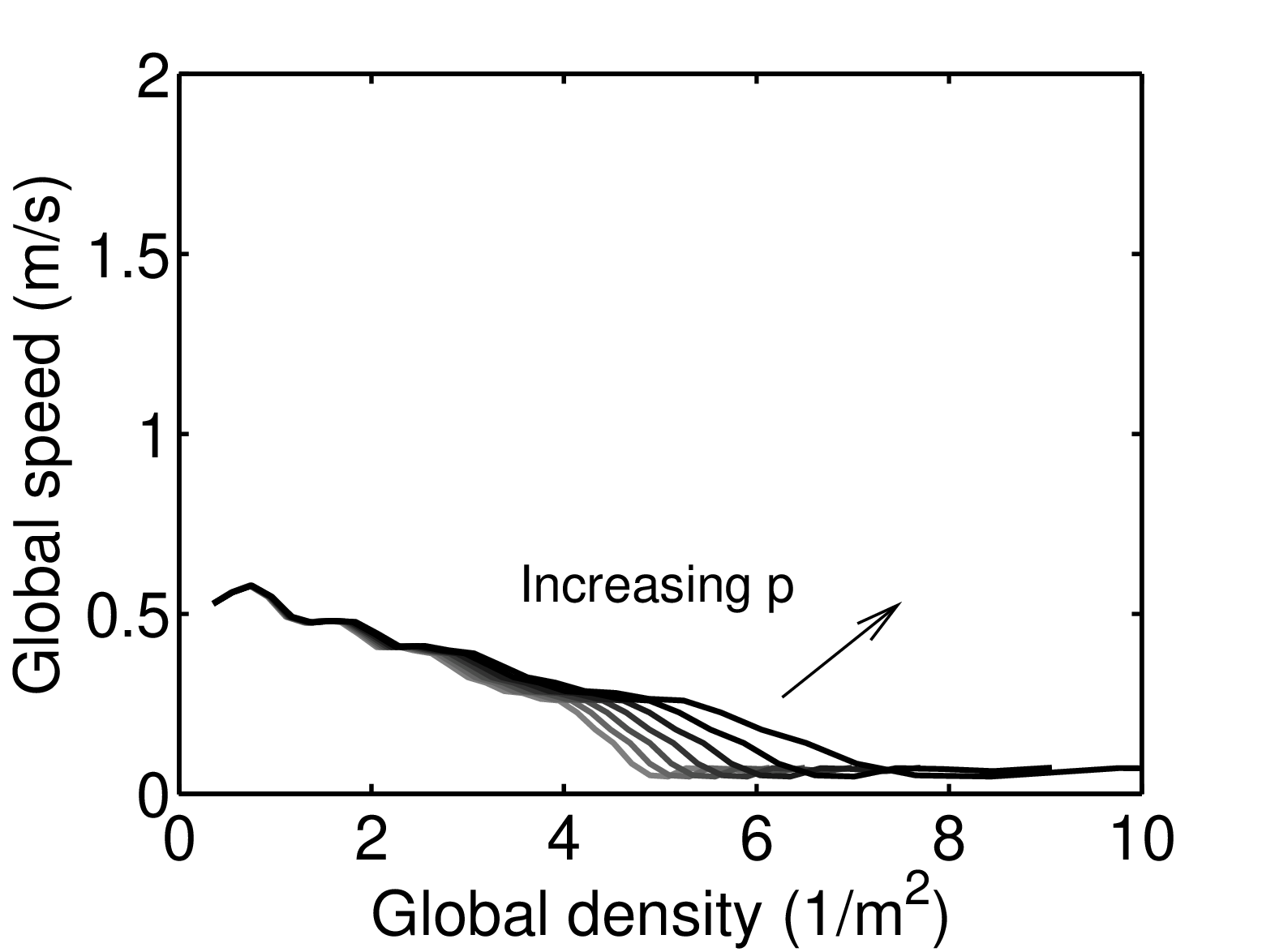}
\includegraphics[width=0.49\textwidth]{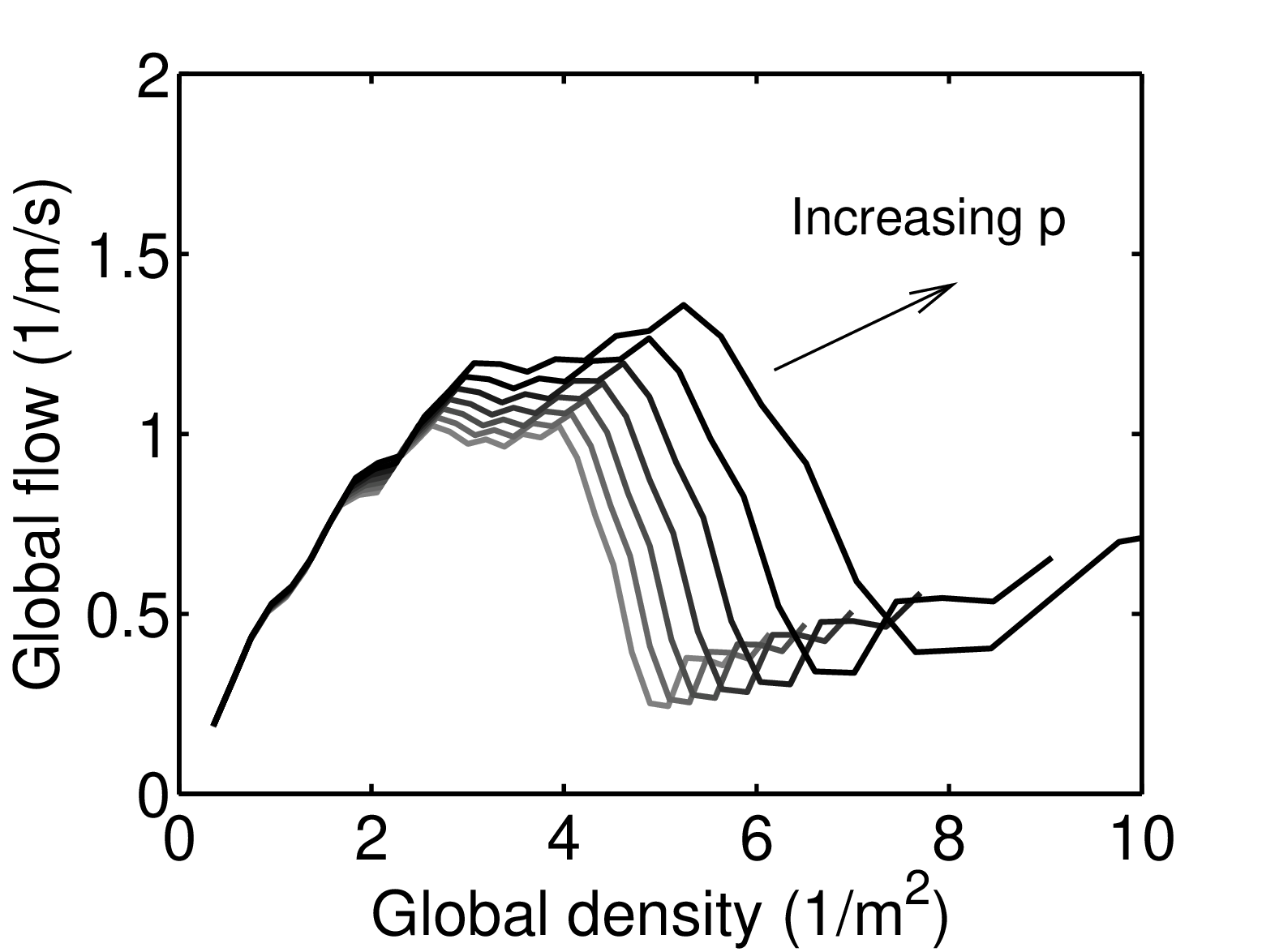}
\caption[]{Relationships between the globally averaged speed and the global density (left) and 
between the globally averaged flow and the global density (right) for various values of 
$p \in  \{0\%, 1\%, 2\%, ..., 6 \%\}$, using formula (\ref{VARRHO}). 
As a larger assumed fraction $p$ of umbrellas implies
a larger number of hidden pedestrians, it is clear that the flow must increase with the
value of $p$.}
\label{p_dependence}
\end{center}
\end{figure}

Moreover, we have determined the globally averaged speed and the 
global flow as a function of the global density for 
different assumed fractions of umbrellas $p$ (see Fig. \ref{p_dependence}).
\par
\begin{figure}[!htbp]
\begin{center}
\includegraphics[width=9cm]{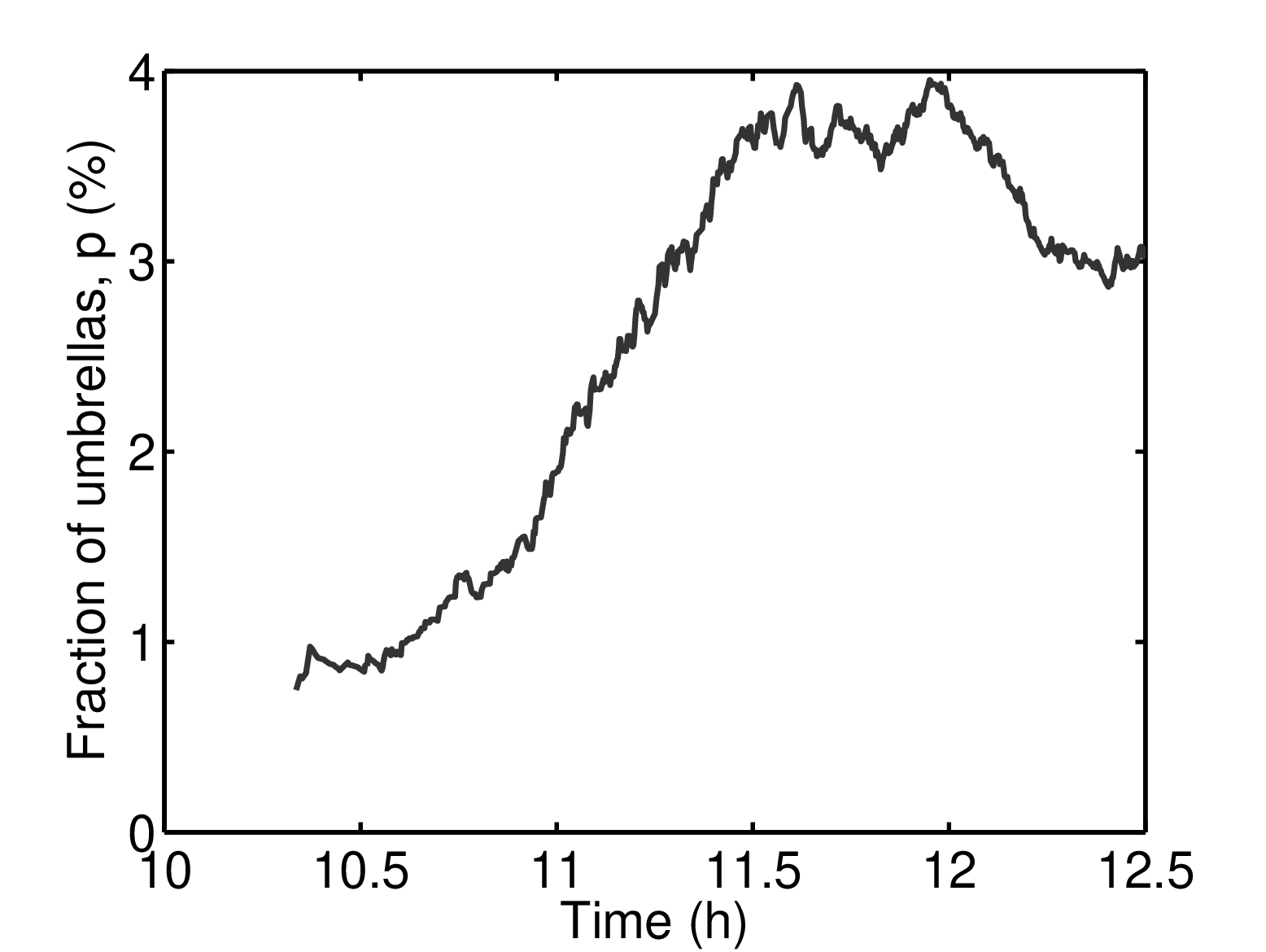}\,
\end{center}
\caption[]{Fraction $p$ of pilgrims carrying umbrellas as a function of time. 
One can see that $p$ is four times larger around noon time, as compared to an hour earlier.}
\label{time_vs_p}
\end{figure}

Let us now address the question, how the {\em actual} fraction of umbrellas can be determined from
the video recordings. For this, we used the fact that the relationship between the local speed and the local
density should not be changed significantly by the fraction $p$ of umbrellas. We, therefore, defined a 
reference relationship $\rho_{\rm ref}(v)$, measured the velocity $v$ and density $\rho$, and
calculated the fraction $p$ of umbrellas that leads to consistent density values. In detail, the procedure
was as follows:
\begin{itemize}
\item As a reference relationship $\rho_{\rm ref}(v)$, we used a fit curve to data of 
the local density as a function of the local speed, determined for $R=1$ and $\gamma = 95$\%.
According to our previous numerical studies, 
the large cutoff value $\gamma$ should eliminate the influence of umbrellas well.
\item At a given time point $t$, we picked $1000$ random locations $\vec{r}_i$.
\item For these locations, we determined the local densities
$\rho'|_{\gamma=0}(\vec{r}_i)$ and the local speeds $v|_{\gamma=0}(\vec{r}_i)$ with
$R=1$m and $\gamma = 0$. The value $\gamma =0$ ignored umbrellas and resulted in
density values $\rho' \le \rho$, but the speeds should be correct. 
\item We then estimated the corrected densities as $\rho(\vec{r}_i) =\rho_{\rm ref}\big(v|_{\gamma=0}(\vec{r}_i)\big)$.  
\item Next, we determined the global density $\varrho(t)$ as average of the corrected local densities
$\rho(\vec{r}_i) $, i.e. $\varrho(t)=\langle\rho(\vec{r}_i)\rangle$. Similarly, we obtained the 
global density $\varrho'(t)$ ignoring umbrellas as average of the densities  $\rho'|_{\gamma=0}(\vec{r}_i)$,
i.e. $\varrho'(t)=\langle \rho'|_{\gamma=0}(\vec{r}_i) \rangle$.
\item Finally, we estimated the fraction of umbrellas via the formula
\begin{equation}
 p(t)=\frac{\varrho(t)-\varrho'(t)}{\varrho^2(t)A_2} \, ,
\end{equation}
which follows from Eq. (\ref{eq_p}). As the radius of most umbrellas was 0.5 meters, we used
the value $A_2=\pi 0.5^2$ $m^2$.
\end{itemize}
The empirically determined fraction $p(t)$ of umbrellas as function of time $t$ is shown
in Fig. \ref{time_vs_p}. It turns out that the fraction of umbrellas increases after 10:30am
and reaches its maximum at noon time, 
when the sunshine is strongest. This shows the plausibility of our procedure, which has also been checked
by manual counts.
\par

\section{Empirical Findings}
\label{fundamental}

\begin{figure}[!htbp]
\begin{center}
\includegraphics[width=0.75\textwidth]{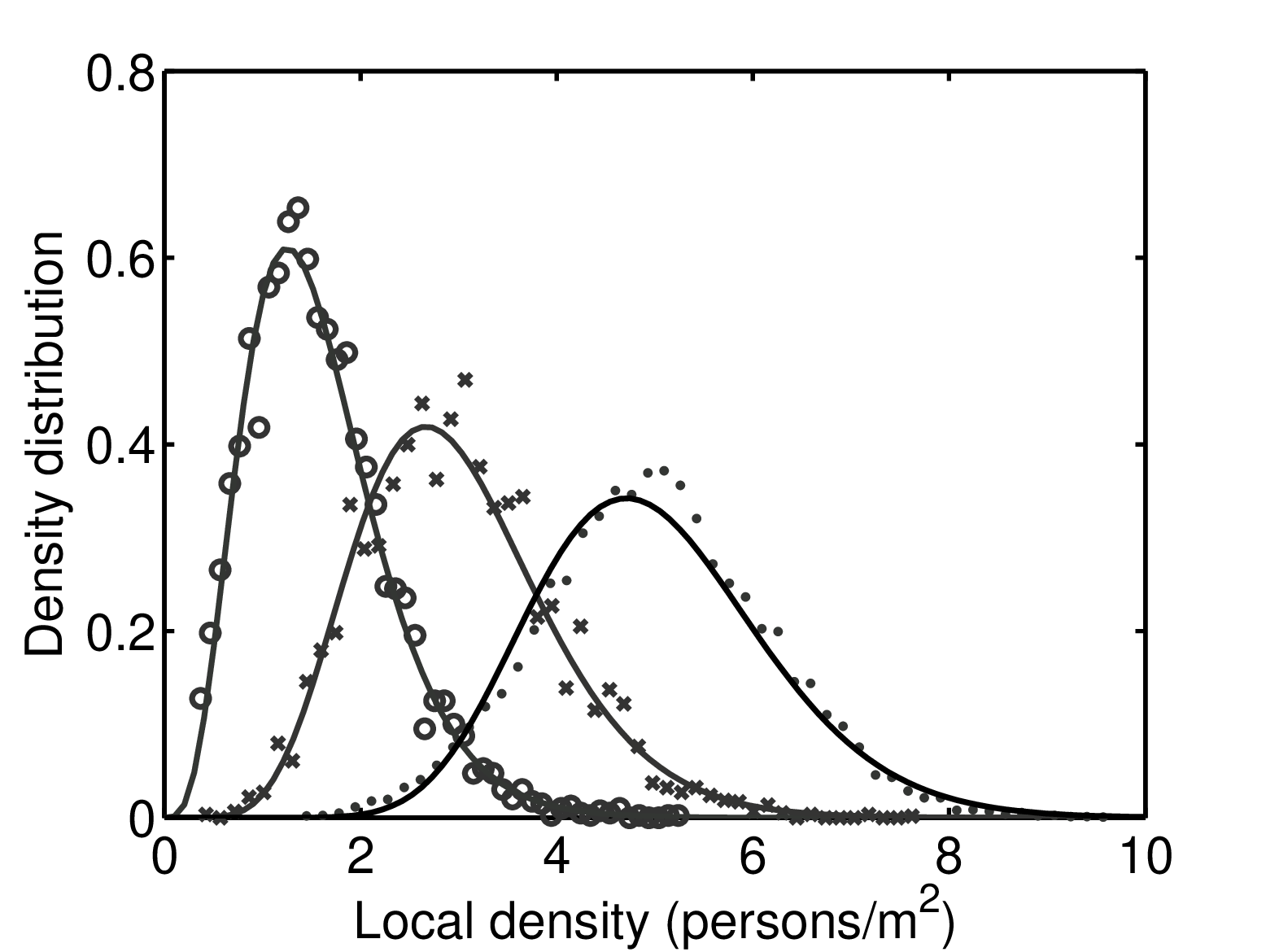}
\end{center}
\caption[]{Distribution of {\it local} densities (with $R=1$) 
for a given {\it average} density (circles: 1.6 persons/m$^2$,
crosses: 3.0 persons/m$^2$, dots: 5.0 persons/m$^2$). Gamma distributions fit the histograms
with 50 bins well (solid lines). After Ref.~\cite{PRE_makkah}}
\label{distribution}
\end{figure}

After the previous description of our video evaluation procedure, let us now start
our analysis of pedestrian recordings with a discussion of density measurements.
Figure \ref{distribution} shows that there is always a large variation of local densities,
and this variation is very important, as the safety in a crowd is not determined by the {\it average} density, but by the {\it maximum} occuring 
density. One can roughly say that the maximum densities are twice as high as the 
average density. Therefore, an average density of 4 persons per square meter should not
be exceeded \cite{Still}.

\subsection{Relationships between Densities, Velocities, and Flows}

\begin{figure}[!htbp]
\begin{center}
\includegraphics[width=0.45\textwidth]{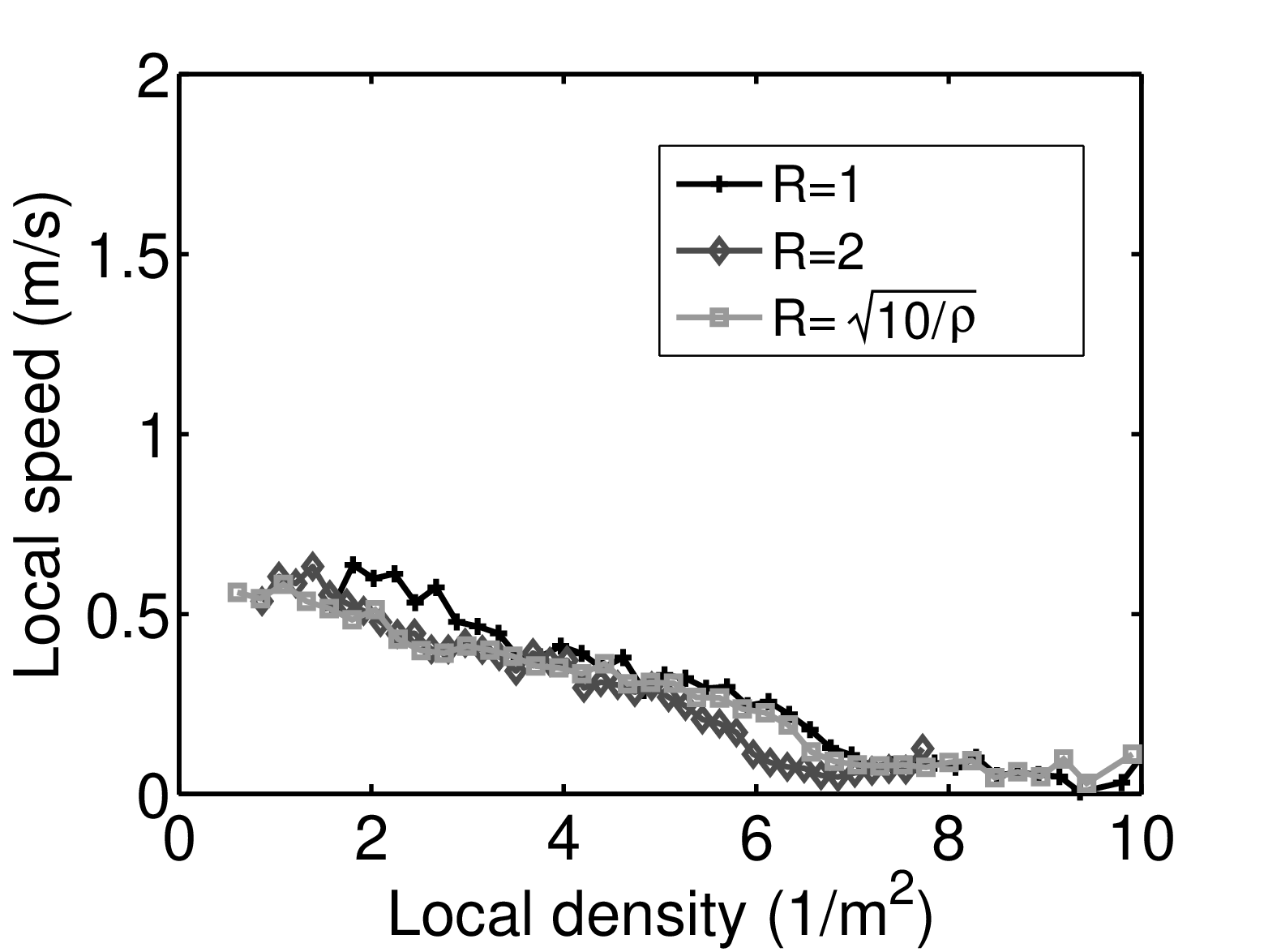}
\includegraphics[width=0.45\textwidth]{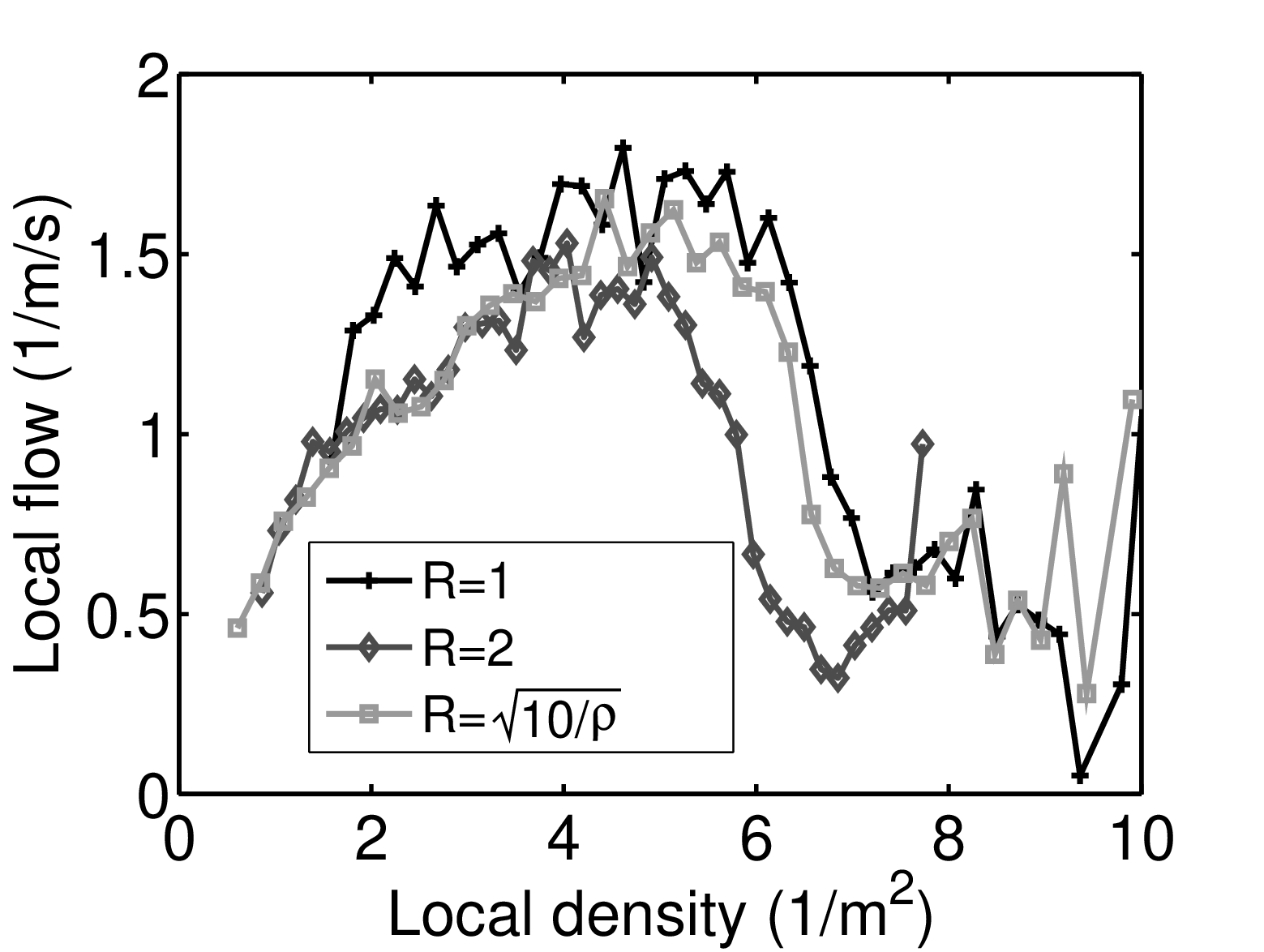}
\end{center}
\caption[]{Average local speed (left) and flow (right) as a function of the local 
density $\rho$. The unexpectedly large range of local densities of up to 10 persons per square meter and higher are in agreement
with manual counts obtained from photographs. Curves are plotted for constant $R = 1$m, $R= 2$m, and 
for the density-dependent specification $R = \sqrt{10/\varrho}$,
corresponding to a constant expected number of pedestrians in the 
area $A_R=\pi R^2$. One can see that smaller values of $R$ tend to imply 
larger flow values, as the averaging is performed
over smaller areas so that extreme values are not averaged out. In other words:
The variation of local densities is smaller the larger the value of $R$. 
As cutoff value we have used $\gamma=0.95$.}
\label{Fig3}
\end{figure}

In this section we are mainly interested in local crowd conditions, as they are relevant for
the criticality in the crowd. Let us first give empirical results for the 
dependence of the locally averaged speed $v(\vec{r},t)= v_t^R(\vec{r})$
and the locally averaged flow $q(\vec{r},t)=q_t^R(\vec{r})$
on the local density $\rho(\vec{r},t)=\rho_t^R(\vec{r})$ (see Fig.~\ref{Fig3}). 
We find that, with larger averaging radius $R$, the
maximum densities and flows are reduced. This is due to the considerable variation
of the local density, speed and flow values. 
In Fig.~\ref{Fig3}, we have therefore also determined the density and flow 
values with a {\em variable} radius $R=\sqrt{10/\varrho}$, in order to average
over a comparable number of pedestrians in all density ranges.
\par
An important observation is the fact that the average local speed at the entrance to the 
Jamarat Bridge does not become zero even at local densities of 10 persons 
per square meter. This, of course, does not mean
that pedestrians never stop, but that stops last for short time periods only (see Ref.~\cite{PRE_makkah}).
The observation of non-vanishing speeds at extreme local densities
is in marked contrast to vehicle traffic, where drivers stop and keep enough distance to avoid collisions.
The consequence of this pedestrian behavior is that their average flows remain finite, with short interruptions only.
However, there is a significant breakdown of the flow by a factor of 3, when the
situation becomes congested. This breakdown implies a serious reduction of the effective capacity,
which causes a further compression of the crowd until the situation becomes critical.
In the worst case, this can lead to crowd accidents. 
\par
In order to avoid a breakdown of pedestrian flows and over-crowding, the
capacity of a pedestrian facility should not be fully utilized. One rather needs sufficient
capacity reserves to guarantee safety with respect to variations in the flows. 
From queuing theory it is known that huge queue lengths
and enormous waiting times result, if the inflow comes close to the flow capacity. 
However, when the waiting times become too long, people become impatient and pushy, 
which further deteriorates the situation.

\subsection{The Fundamental Diagram and its 
Comparison with Other Measurements}

\begin{figure}[!htbp]
\begin{center}
\includegraphics[width=0.45\textwidth]{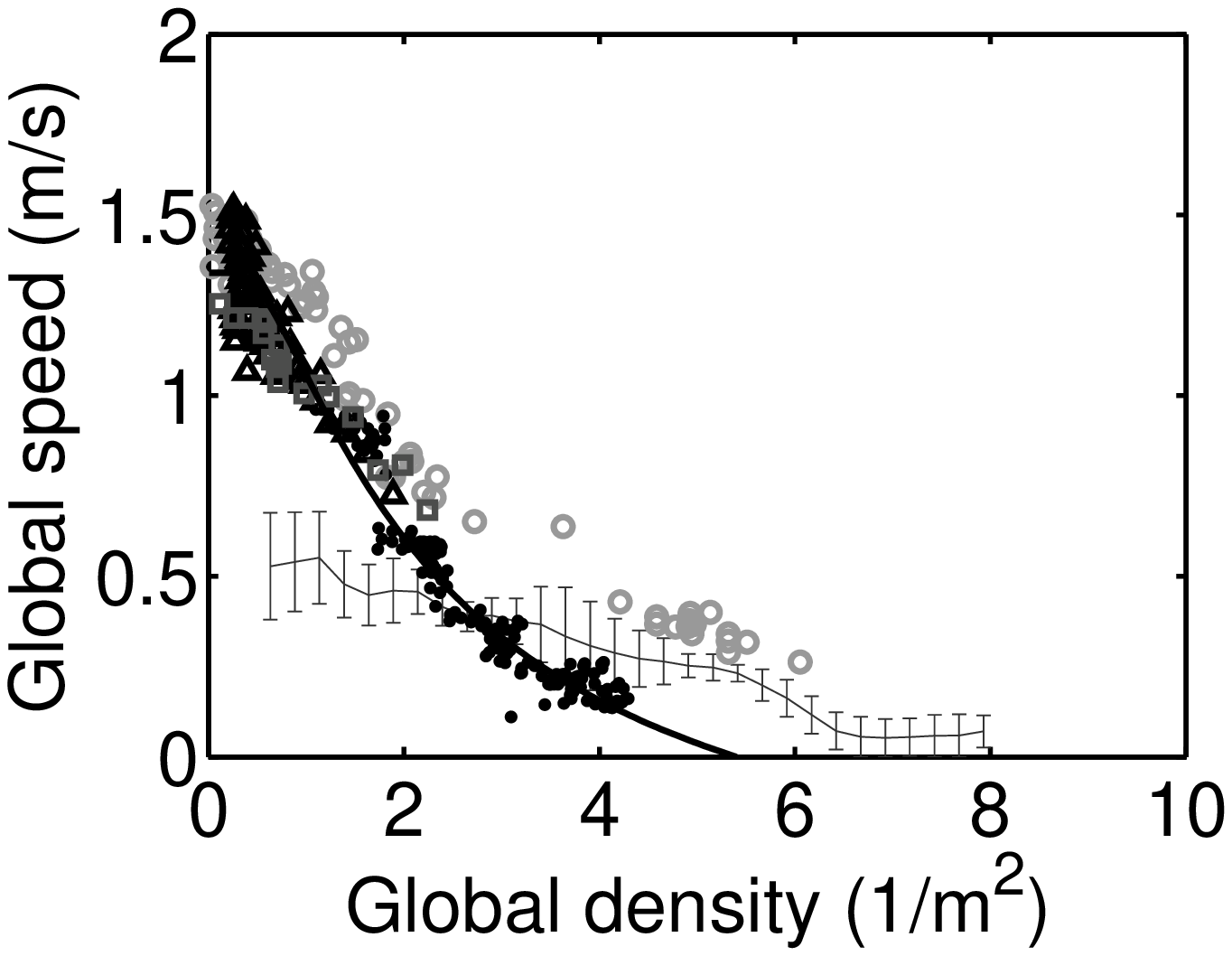}\,
\includegraphics[width=0.45\textwidth]{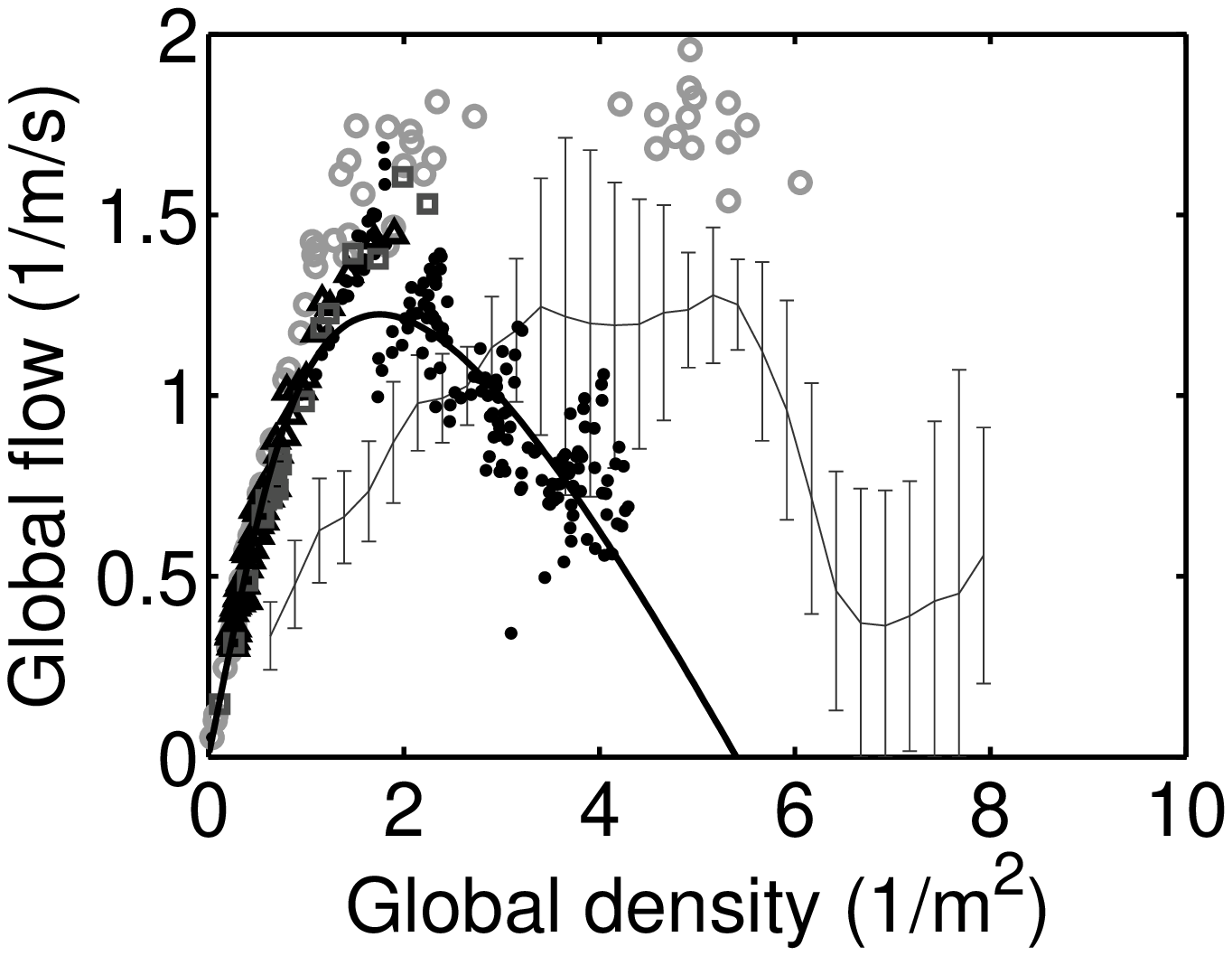}
\end{center}
\caption[]{Global speed $V$ (left) and global flow $Q$ (right) as a function of the global density $\varrho$
in the video-recorded area. Our own data  were determined
using the time-dependent values $p(t)$ depicted in Fig.~\ref{time_vs_p} in order
to correct for umbrellas. The average speeds and flows, obtained by averaging all 
data points for the same density, are represented by solid lines with error bars corresponding to
one standard deviation (Note that most publications on the fundamental diagram of pedestrian flows
do not provide error bars which must be criticized for neglecting the considerable variety of flow values that is also
known for vehicular traffic). While variations of the average speeds appear to be 
relatively low, the flow values vary significantly, which must be taken into account
by 30--40\% safety margins in any capacity assessment.
Symbols correspond to the empirical data 
of Mori and Tsukaguchi \cite{Mori} (circles), Polus {\it et al.} \cite{Polus}  (squares), {Fruin {\it et al.} \cite{Fruin2} (triangles),}
 and Seyfried {\it et al.} \cite{Seyfried} (dots). 
The solid fit curve is from Weidmann \cite{Weidmann}.
Note that the data by Mori and Tsukaguchi were not averaged over a large area
and, therefore, rather represent local measurements. Note that the global
flows determined from our measurement were bounded by the capacity of the stoning ritual
at the pillars of the Jamarat Bridge, i.e. the maximum global flows at 3 to 6 persons 
per square meter could potentially be higher (as the local flow-density curves suggest).
It should be stressed that the measurements here are {\em global}, i.e. averaged over the whole measurement area. Due to the large variability of local densities (see Fig. \ref{distribution}), this gives 
significantly lower densities and flows as compared to the similarly looking curves for {\it local} measurements presented in Fig. \ref{Fig3}, Fig. 1 of Ref. \cite{PRE_makkah}, and Mori and Tsukaguchi \cite{Mori}.
}
\label{average}
\end{figure}

Note that the actual capacity of a pedestrian facility can {\it not} be determined from 
Fig. \ref{Fig3}, as the maximum of local flows is much higher than the maximum of
the {\em average} flows, i.e. the capacity is significantly {\it below} the maximum of this curve! 
The relationship between the {\it average} flow as a function of the {\it average}
density is, therefore, displayed in Fig. \ref{average}. This curve is called
the {\em ``fundamental diagram''}. Its maximum flow is often used for capacity assessments.
\par
In order to compare our measurements with previous studies, we have digitized corresponding
data published in the pedestrian-related literature. Figure \ref{average} presents a
comparison with various measurements. It turns out that the maximum average (``global'') densities are
higher and the average speeds are lower than the ones reported in many publications. 
Obviously, it is important to identify the reasons for this.  A closer analysis shows that
most data displayed in Fig. \ref{average} are for European or North-American countries, while
Mori and Tsukaguchi's measurement was carried out in an Asian country, where people are smaller.
Therefore, {\em the body size distribution has a dramatic influence on the velocity-density relationship
and the fundamental diagram} \cite{Pheasant,Teknomo,Pauls}. It is consistent with observations
that the average ``diameter'' of pilgrims is smaller than that of European or American citizens, which
explains the higher maximum densities observed in our study. 
\par
\begin{figure}[htbp]
\begin{center}
\includegraphics[width=0.45\textwidth]{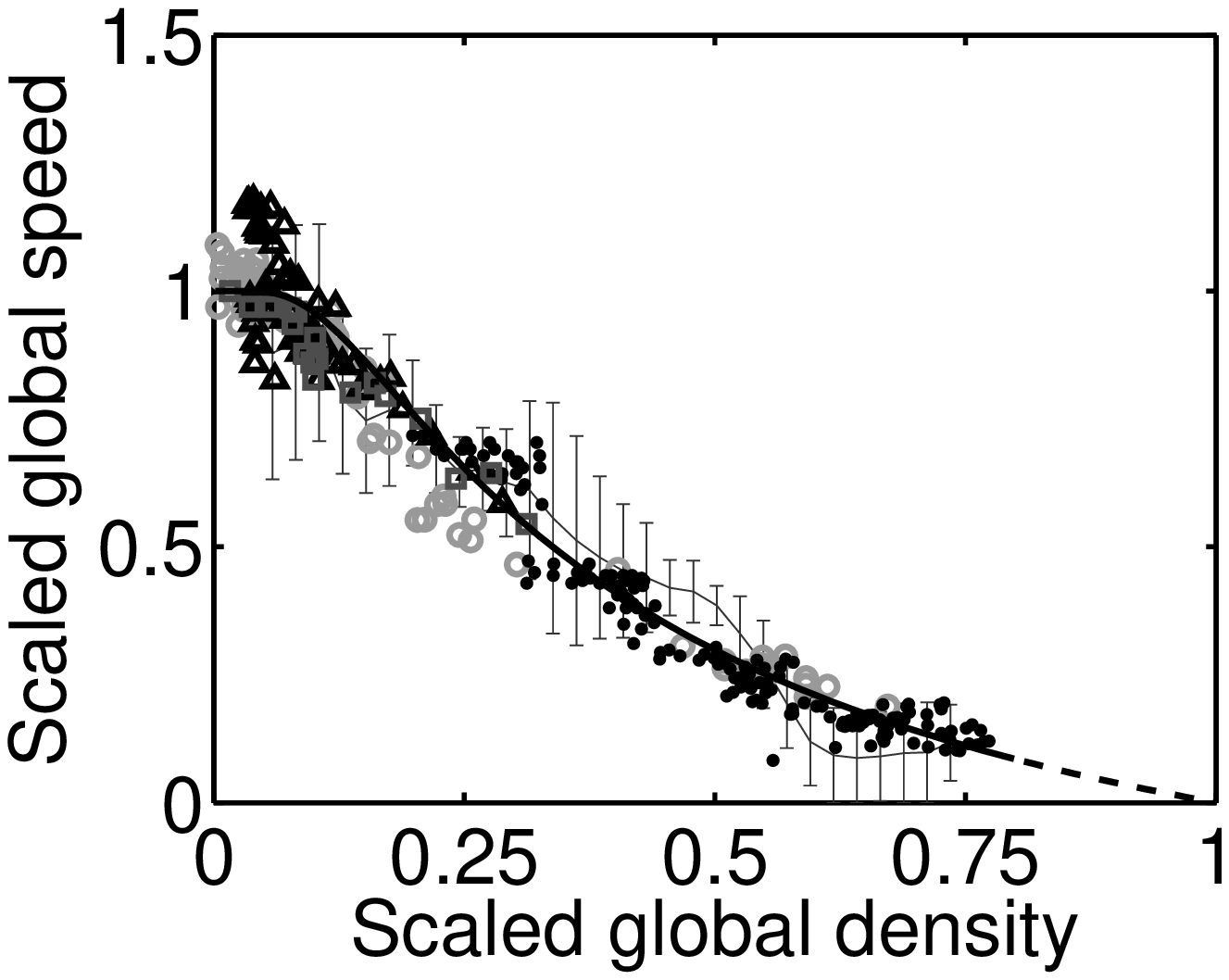}\,
\includegraphics[width=0.45\textwidth]{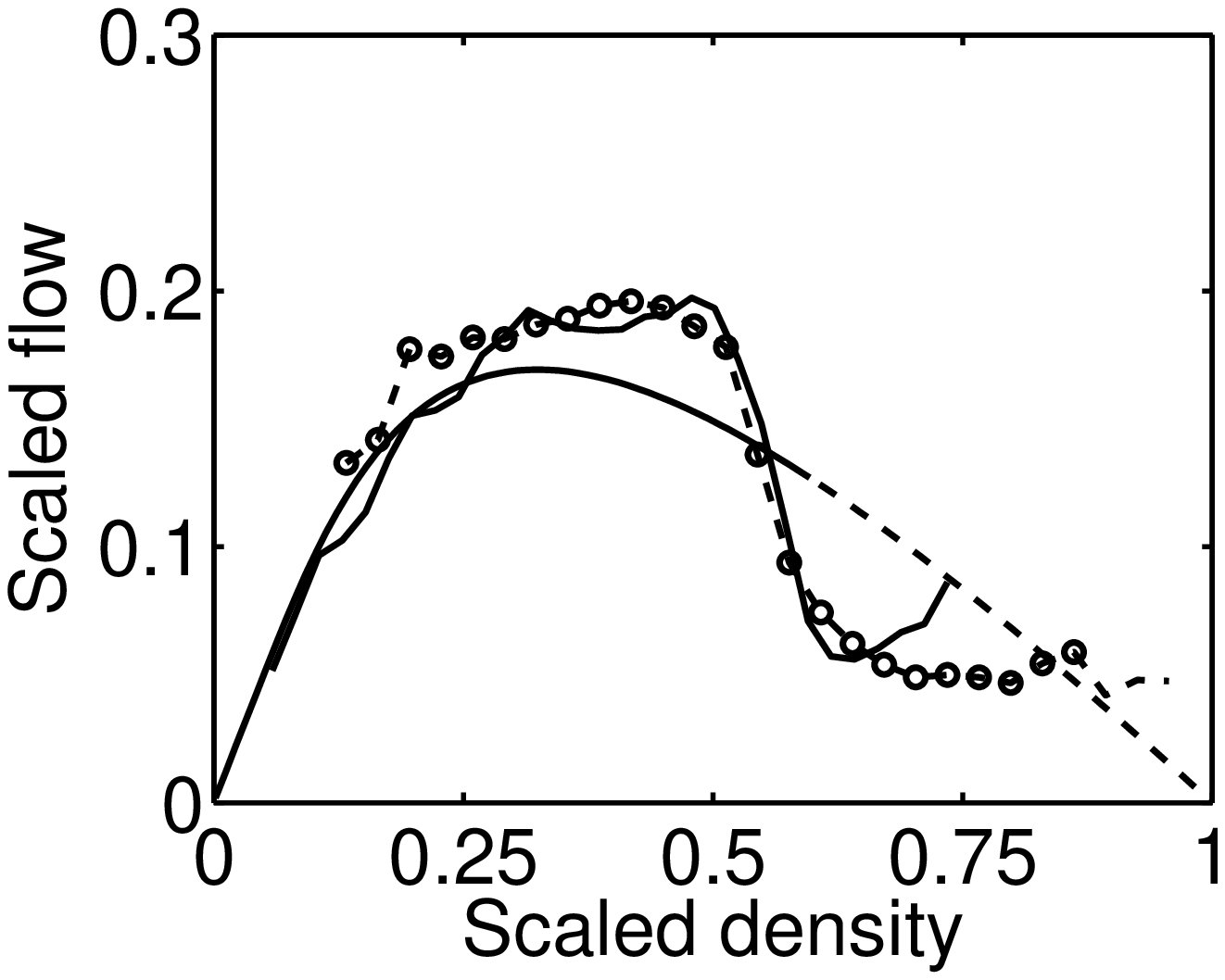}\,
\end{center}
\caption[]{Left: Comparison of scaled empirical measurements of the speed-density relationship 
with Weidmann's curve (\ref{Wei}) (dashed line, see main text for details): The fit parameters for 
Mori's and Tsukaguchi's data \cite{Mori} (circles) are $V_0 = 1.40$ m/s and $\rho_{\rm max} = 9.00$ persons/m$^2$, for Polus {\it et al.} \cite{Polus}  (squares) they are $V_0 = 1.25$ m/s and $\rho_{\rm max} = 7.18$ persons/m$^2$, 
for Fruin {\it et al.} \cite{Fruin2} (triangles) they are $V_0 = 1.30$ m/s and $\rho_{\rm max} = 6.60$ persons/m$^2$, 
for Seyfried {\it et al.} \cite{Seyfried} (dots) they are $V_0 = 1.34$ m/s and $\rho_{\rm max} = 5.55$ persons/m$^2$,
and for our data (thin solid line with error bars) they are $V_0 = 0.60$ m/s and $\rho_{\rm max} = 10.79$ persons/m$^2$.
The rescaling was done by first identifying $V_0$, either directly from the data or by extrapolating the speed data towards zero density.
Then, a least-square fit was performed to find the value of $\rho_{\rm max}$ giving the best match of Eq. (\ref{Wei}). 
Finally, the densities were scaled by $\rho_{\rm max}$, while the speeds were scaled by $V_0$. 
It can be seen that scaled speed-density data from different places in the world are reasonably compatible.
Right: The compatibility of flow-density data is less obvious, particularly at extreme densities: The smooth curve was derived from Weidmann's speed-density relation (\ref{Wei}), while the other solid curve corresponds to our scaled global flow as a function of the scaled global density. The dashed line (highlighted by circles) represents the scaled local flow (for $\gamma = 0.8$) as a function of the scaled local density with $V_0 = 0.86$ meters per second and $\rho_{\rm max} = 10.57$ persons per square meter. Therefore, it must be stressed that flow-density relations should not be derived from
speed-density fits, but separately fitted to empirical flow data.
}\label{NEW}
\end{figure}

It would be desireable, if measurements for different countries could be mathematically transformed to a universal curve, and if country-specific diagrams could be derived from it by determining a few parameters only. For example, one could think of using 
a density definition based on projected body areas, as proposed in Ref. ~\cite{Predtechenskii}. This approach, however, is limited
for the following reasons:
\begin{enumerate}
\item Densities still vary considerably, when the projected body areas have already reached 100\% spatial coverage.
\item The coverage of two-dimensional space by projected body areas is hard to measure, as it requires perpendicular recording, which is possible only in a small number of cases.
\item Favorable distances and speeds depend a lot on the cultural settings. For example, Ref. \cite{RichardWiseman} shows that there are tremendous differences in walking speeds for different populations: The fastest average walking speed (Singapore) is three times higher than the slowest  one (Malawi). During the Hajj, the low average speed at small densities  is due to the
fact that pilgrims tend to walk together in groups, and a considerable number of group members is 50 years or older.
\end{enumerate}
Simple transformations can certainly not account for the varity of cultural factors, but a rough approximation may be made as follows:
Determine the average speed at very low densities by measurement or extrapolation, and identify the value with $V_0$. Calculate
the expected speed as a function of the density $\rho = x\rho_{\rm max}$ as
\begin{equation}
V(x\rho_{\rm max}) = V_0 F(x)
\label{One}
\end{equation}
or the flow as a function of the density as
\begin{equation}
Q(x\rho_{\rm max}) = x\rho_{\rm max} V_0 F(x) \, ,
\label{Two}
\end{equation}
where
\begin{equation}
F(x) = 1 - \exp \left[ -0.35 \left(\frac{1}{x} - 1 \right)\right]
\label{Wei}
\end{equation}
corresponds to Weidmann's curve (\ref{WEID}). As the maximum density $\rho_{\rm max}$ is hardly measurable directly,
it is treated as a fit parameter somewhere in the range between 4 and 12 persons per square meeter. and determined by minimizing the deviation of (\ref{One}) or (\ref{Two}) from empirical measurements for the country or event of interest. Correspondingly scaled results are shown in Fig. \ref{NEW}. The same procedure could be applied to improved standardized functions $F(x)$, which take into accunt particularities at 
extreme densities.

\section{Warning Signs of Critical Crowd Conditions} \label{critic}

\subsection{Transition to stop-and-go waves} \label{sec_stopandgo}

\begin{figure}[htbp] 
\unitlength1cm
\begin{center}
\begin{picture}(11,15)
\put(0,9.6){\includegraphics[width=0.865\textwidth]{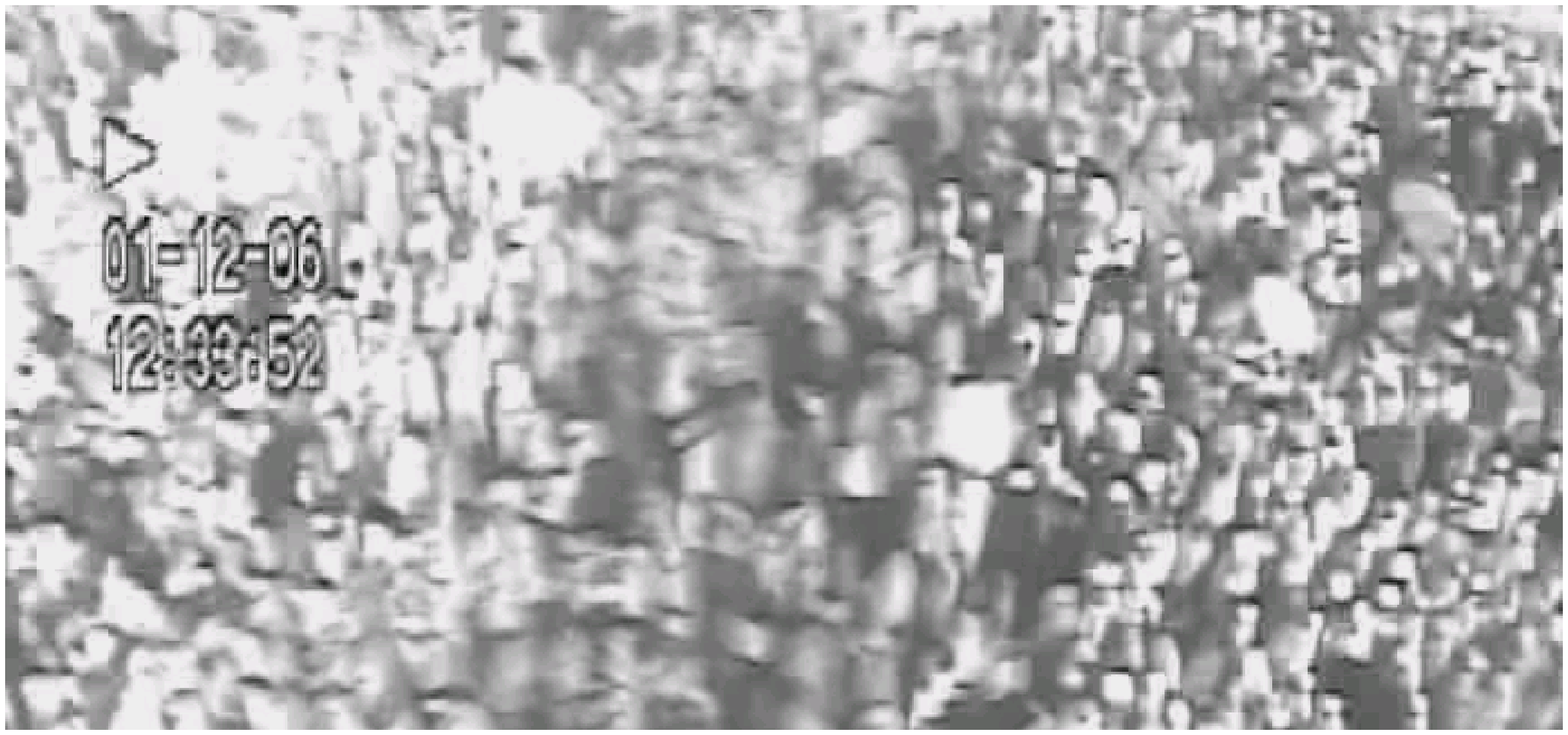}}
\put(-1.2,3.7){\includegraphics[width=1\textwidth]{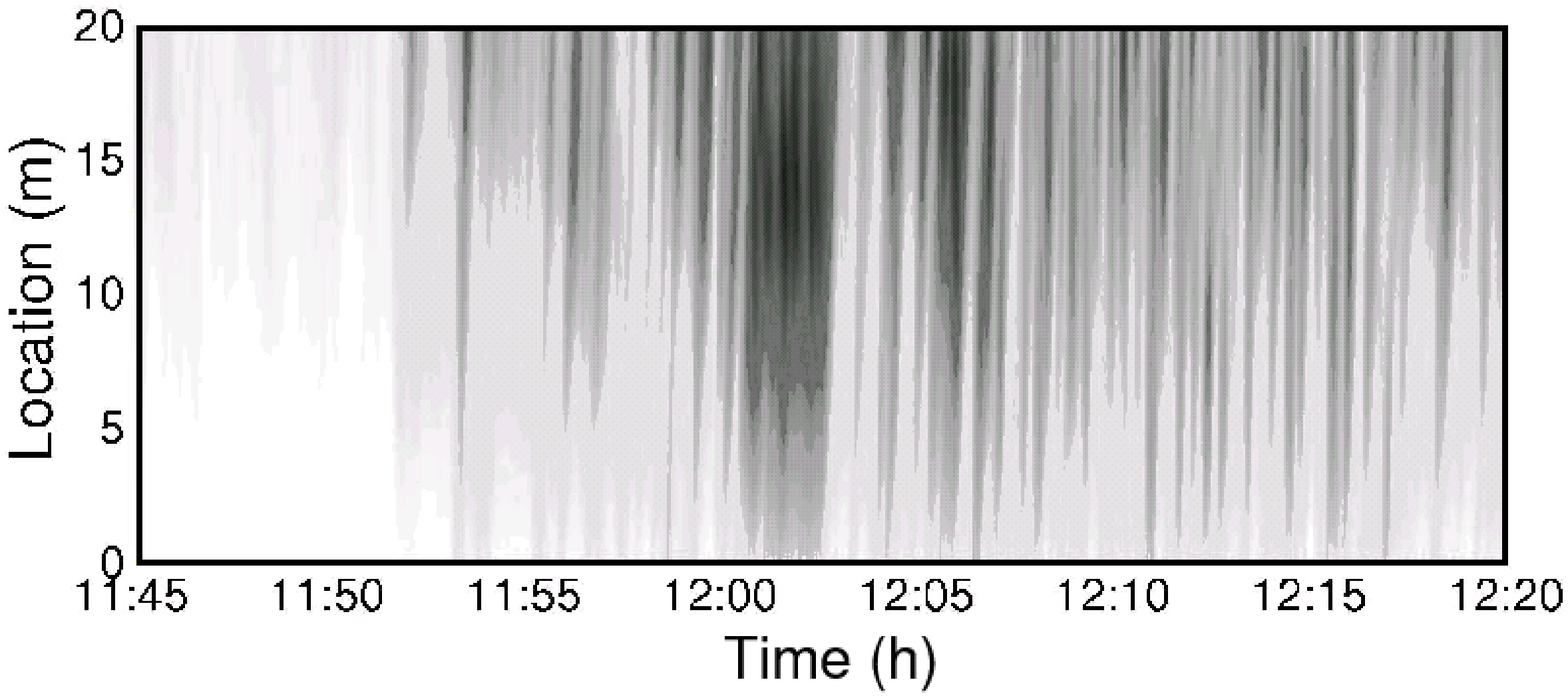}}
\put(4.07,-1.9){\includegraphics[width=0.6\textwidth]{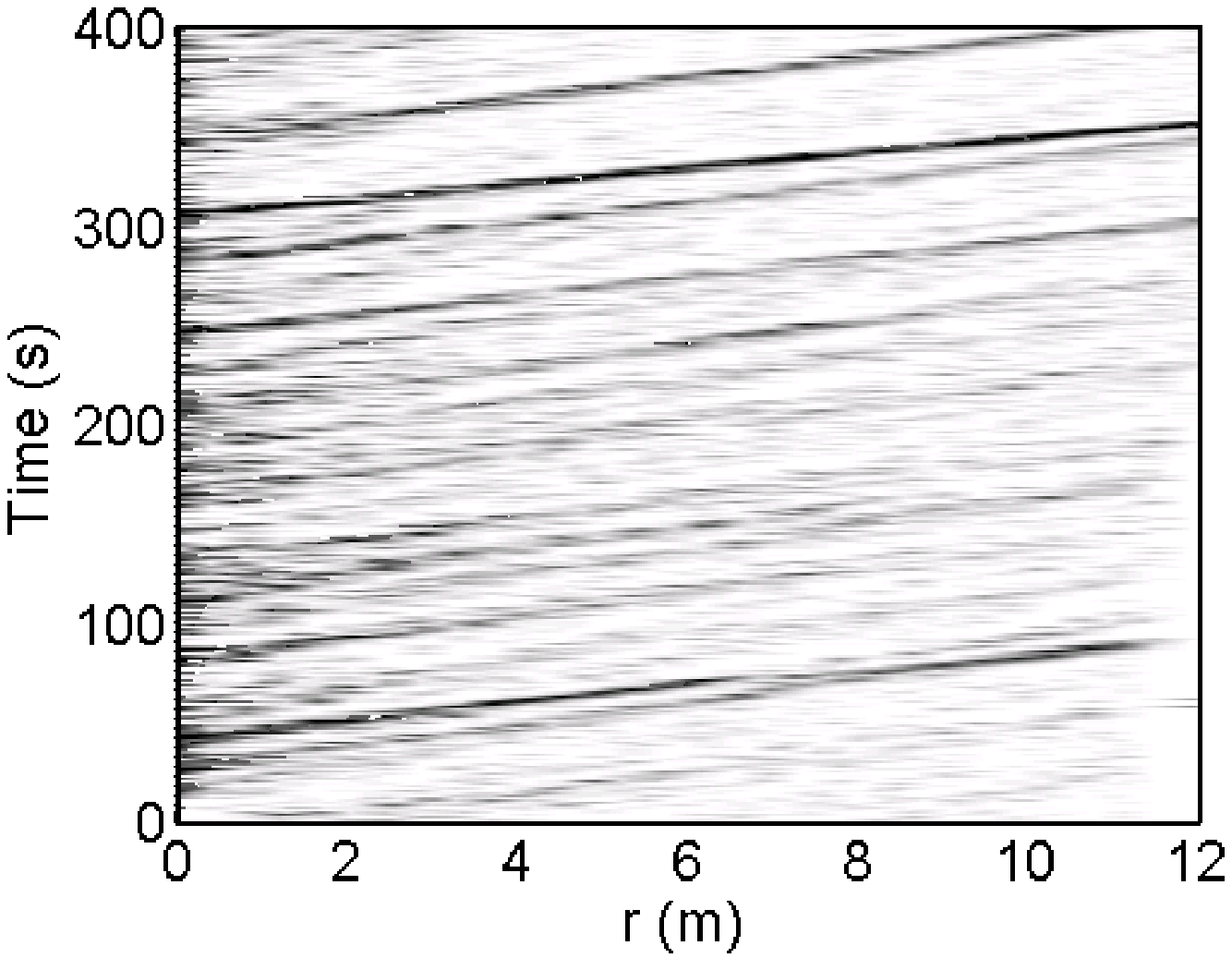}}
\put(-0.3,-1.4){\includegraphics[width=0.4\textwidth,angle=90]{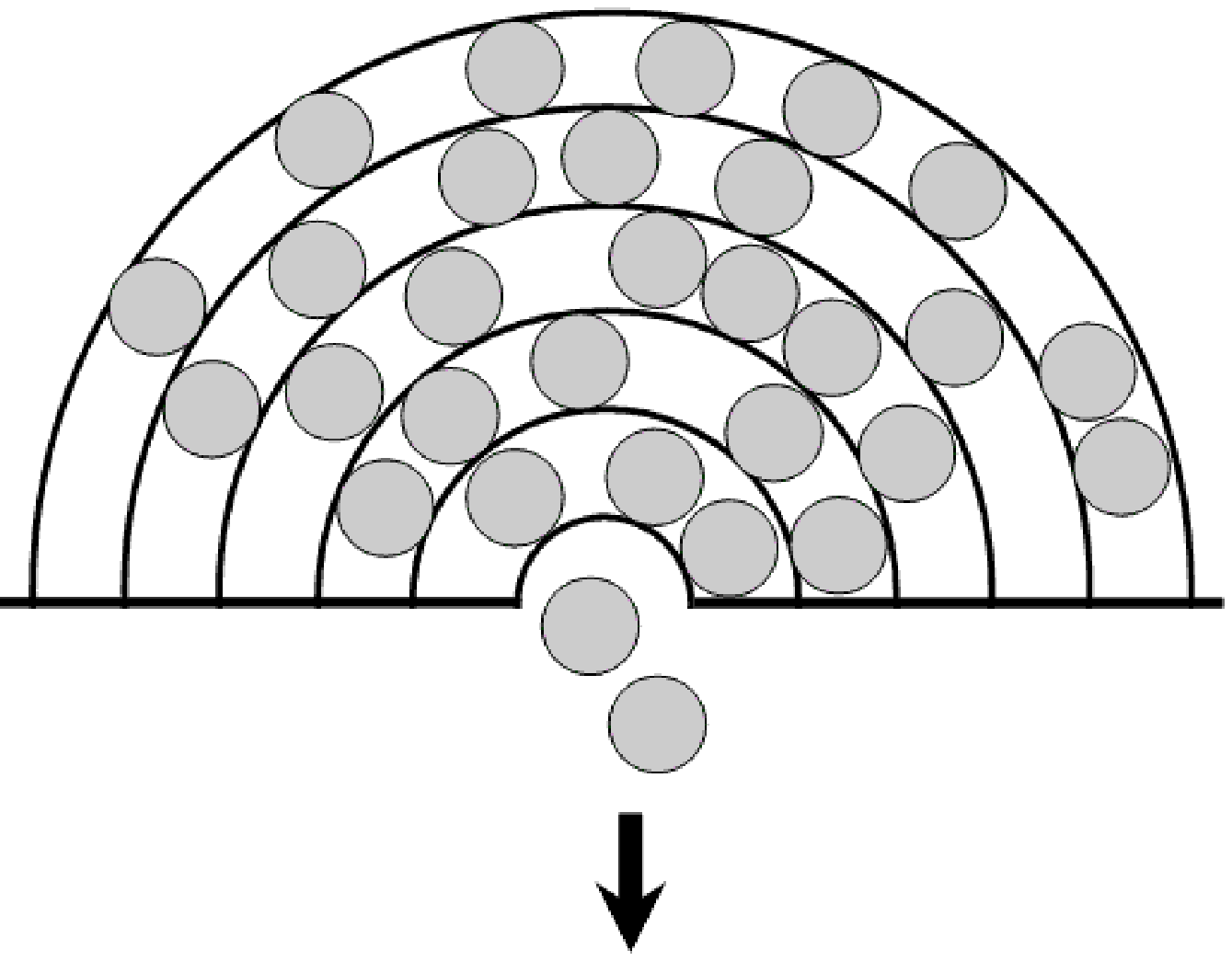}}
\end{picture}
\end{center}

\vspace{1.3cm}

\caption[]{Top: Long-term photograph showing stop-and-go waves in a densely packed street.
While stopped people depicted relatively sharp, people moving from right to left 
have a fuzzy appearance. Note that gaps propagate from left to right. Middle:
Empirically observed stop-and-go waves in front of the entrance to the Jamarat Bridge on 
January 12, 2006 (after \cite{PRE_makkah}), where pilgrims moved from left to right. Light areas
correspond to phases of motion, dark colors to stop phases. The ``location''  coordinate
represents the  distance to the beginning of the narrowing, i.e. to the cross section of reduced width. Bottom left:
Illustration of the ``shell model'' (see Ref. \cite{BottleneckPRL}), in particular of situations where several pedestrians compete
for the same gap, which causes coordination problems. Bottom right: Simulation results of
the shell model. The observed stop-and-go waves 
result from the alternation of forward pedestrian motion and backward gap propagation.}
\label{stop}
\end{figure}

Our study of pilgrim flows has also revealed dynamical phenomena. As one of the videos provided on the supplementary 
webpage \cite{supplement} shows, at high crowd densities there was
a sudden transition from  laminar flows to stop-and-go waves upstream of 
the 44 meter wide entrance to the Jamarat Bridge \cite{PRE_makkah}.
These propagated over distances of more than 30 meters (see Fig. \ref{stop}). The sudden transition
was related to a significant drop of the flow, i.e. with the onset of congestion \cite{PRE_makkah}.
Once the stop-and-go waves set in, they persisted over more than 20 minutes.
\par
This phenomenon can be reproduced by a recent model based on two continuity equations, one
for forward pedestrian motion and another one for backward gap propagation \cite{BottleneckPRL}. 
The model was derived from a ``shell model'' (see Fig. \ref{stop}) and describes 
very well the observed alternation between backward gap propagation and forward pedestrian motion.

\subsection{Transition to ``crowd turbulence''}

\begin{figure}[htbp] 
\begin{center}
\includegraphics[width=10.5cm]{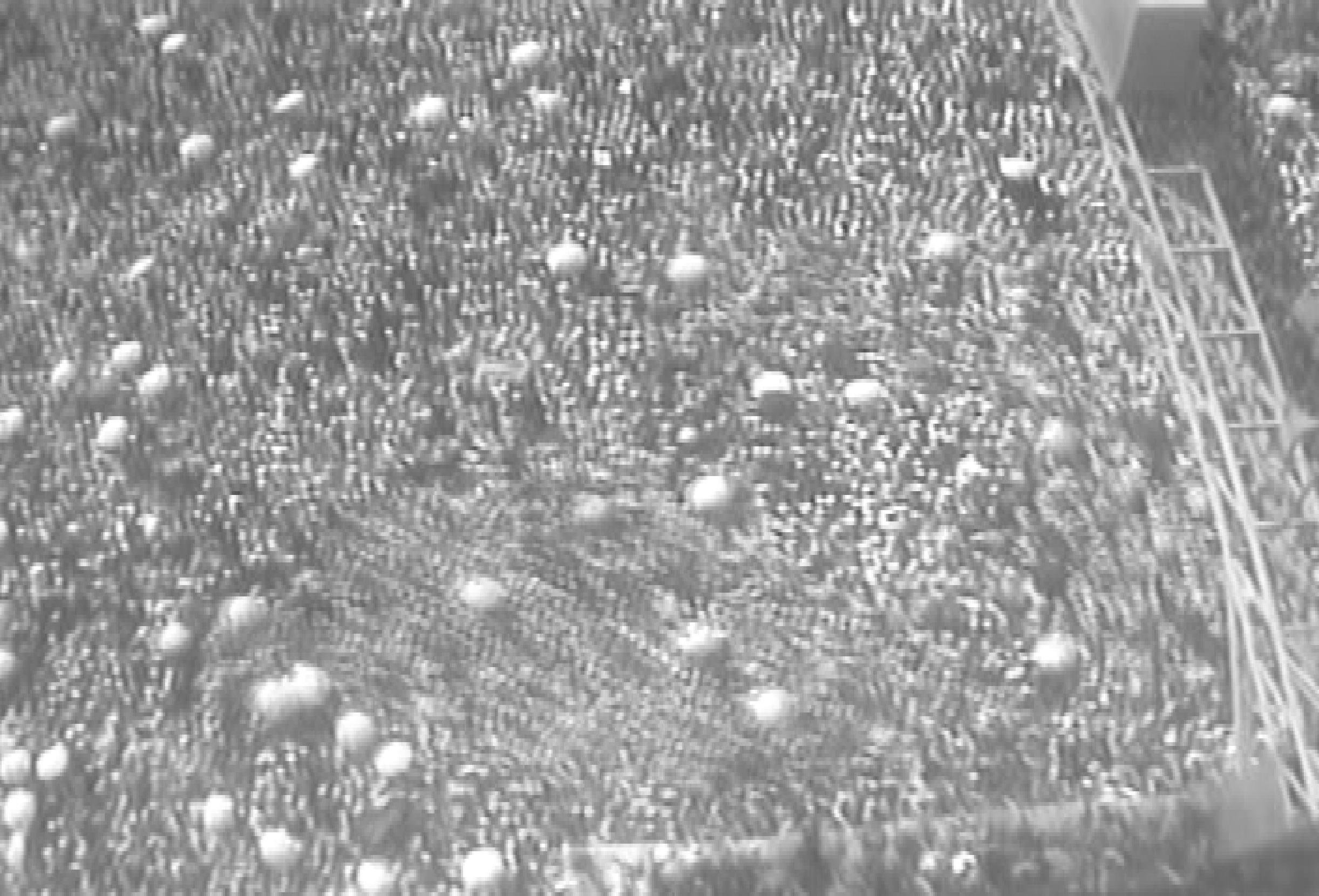}\,
\end{center}
\caption[]{Long-term photograph of the phenomenon of ``crowd turbulence''. In the fuzzy area of this picture,
people are moved into all possible directions by physical forces building up in extremely dense crowds.}
\label{turbul}
\end{figure}

After the occurence of the stop-and-go waves and the related breakdown of the pedestrian flow,  
the density reached even higher values and 
the video recordings showed a sudden transition from stop-and-go waves
to {\em irregular} flows  (see Fig.~\ref{turbul}). These irregular flows were characterized
by random, unintended displacements into all possible directions, which pushed people 
around (see Ref. \cite{PRE_makkah} for sample trajectories). This ``crowd turbulence'' caused some individuals to stumble. 
As the people behind were moved by the crowd as well and could not stop, fallen 
individuals were trampled, if they did not get back on their feet quickly enough. 
Tragically, the area of trampled people grew more and more in the course of time, 
as the fallen pilgrims became obstacles for others \cite{PRE_makkah}. The result was one of the
biggest crowd disasters in the history of the pilgrimage.
\par
How can we understand this transition to irregular crowd motion?
A closer look at video recordings of the crowd reveals that, 
at the time when the phenomenon of ``crowd turbulence'' occured, people were so densely packed that they  
were moved involuntarily by the crowd. Recently, a computer simulation of this situation was made
\cite{turbulenceModel}. It managed to reproduce the main observations with an extended social force model, 
assuming that people would try to gain space when the density becomes very high.

\subsection{Measuring Critical Crowd Conditions}

\begin{figure}[!htbp]
\begin{center}
\includegraphics[width=10.5cm]{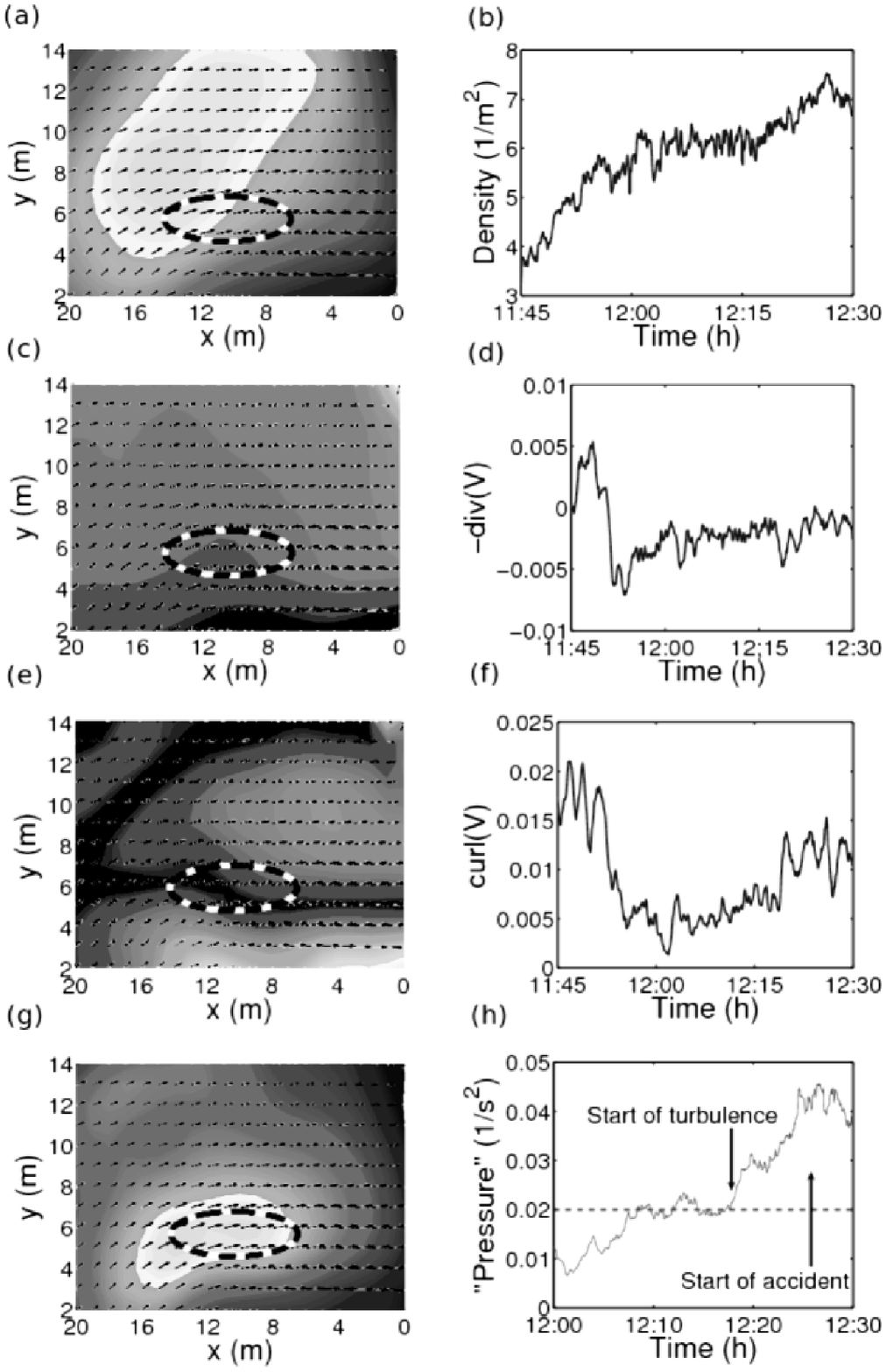}\,
\end{center}
\caption[]{Illustrations of (a, b) the density, (c, d) the negative of the divergence of the velocity field, (e, f) the curl of the velocity field, and (g, h) the ``crowd pressure''. The figures on the left show the quantities as functions of space, while the figures on the right
show the same quantities as functions of time. A color version of these figures can be found on the supplementary web page \cite{supplement}. The spatial plots are time-averages over the period from 11:45 to 12:30 on January 12, 2006. The arrows show the velocity field, obtained by averaging over the time interval from 11:45 to 12:00.  Brighter values correspond to higher values. The dashed ellipse indicates the area where the crowd accident on January 12, 2006, started.}
\label{spatial_and_temporal_plots}
\end{figure}
For a successful crowd management and control, it is important to know where and
when critical situations are likely to occur, although it must be always kept in mind that
not all hazards to the crowd can be reliably detected.
Visual inspection of surveillance cameras
is not very suited to identify critical crowd conditions: 
While the average density rarely exceeds
values of 6 persons per square meter, the local densities can vary considerably 
(see Fig.~\ref{distribution}). Moreover, the contour plot of the density and its temporal evolution do not give a precise 
answer, where and when the situation becomes critical (see Figs. \ref{spatial_and_temporal_plots} a+b): At a certain 
time, the density is high almost everywhere in the recorded area. However, 
continuous alarms as given by warning systems in the past are not very useful, since security forces 
cannot be everywhere. Therefore, it is important to focus their attention and activity on specific areas. 
\par
Also the analysis of the velocity field is not particularly suited to 
identify critical areas and times. Figures \ref{spatial_and_temporal_plots} c+d show the 
negative divergence of the time-averaged velocity field times the 
density which, according to the continuity equation, describes 
the expected density increase along the trajectory of a pedestrian.\footnote{From the continuity
equation $\partial \rho/\partial t + \vec{\nabla} (\rho\vec{v}) = 0$ follows
$(d\rho/dt)/\rho = -\vec{\nabla} \vec{v}$, where $d\rho/dt = \partial \rho/\partial t + \vec{v}\cdot\vec{\nabla}\rho$
represents the density change in time in the moving system, i.e. along an average trajectory.}
Moreover, Figs. \ref{spatial_and_temporal_plots}e+f show the magnitude of the curl of the time-averaged velocity field measuring the vorticity. Again, these figures give no indication of the time
and location of the sad crowd accident.
\par
The decisive variable quantifying the hazard to the crowd is rather 
the variance of speeds, multiplied by 
the density. We call this quantity the 
``(crowd) pressure''.\footnote{This ``gas-kinetic'' definition of the pressure is to be distinguished from the 
mechanical pressure experienced in a crowd. However, a monotonously increasing
functional relationship between both kinds of pressure is likely, at least when averaging over
force fluctuations.} It allows one to identify critical locations (see Fig. \ref{spatial_and_temporal_plots}g) 
and times (see Fig. \ref{spatial_and_temporal_plots}h). There are even advance warning signs of critical crowd conditions:
The crowd accident on January 12, 2006 started about 10 minutes 
after turbulent crowd motion set in, which happened when
the ``pressure'' exceeded the ($R$-dependent) value of 0.02/s$^2$ \cite{PRE_makkah}
(see Fig. \ref{spatial_and_temporal_plots}h). Moreover, it occured more than 30 minutes after the average
flow dropped below the (also $R$-dependent) threshold of 0.8 pilgrims per meter and second,
which can be identified as well by watching out for stop-and-go waves in accelerated 
surveillance videos (played fast-forward) \cite{PRE_makkah}. 
Such advance warning signs of critical crowd conditions 
could be evaluated on-line by a video analysis system. 
This would allow one to gain time for corrective measures like flow control, re-routing of people,
pressure relief strategies, or the separation of crowds into blocks to stop the propagation
of shockwaves. Such anticipative crowd control could certainly increase the level of safety during 
future mass events. 

\section{Summary and Discussion}\label{summary}

While most previous measurements of the fundamental diagram for pedestrian flows
have been restricted to densities upto 4 persons
per square meter, we have measured local densities of 10 persons per square meter and more
on the 12th day of the Muslim pilgrimage in 1426H close to the entrance ramp of the Jamarat Bridge.
The occurence of such high densities relates to the body size distribution of pilgrims and
implies the possibility of unexpectedly high flow values.
Even at densities up to 10 persons per square meter, the average motion of the crowd is not entirely stopped. This can lead 
to over-critical compressions. However, the density is not the optimum indicator of critical
crowd conditions. The ``pressure'', defined as the density times the variance of velocities, 
provides better and more specific information about critical areas and times.
\par
Our results have been obtained with a newly developed, powerful video analysis method for pedestrians, which is well applicable even to dense pedestrian crowds. The automated evaluation delivers pedestrian counts in different directions of motion and measurements of local densities, speeds, flows, and pressures. This information can be used for an improved surveillance of the crowd and
the identification of critical crowd conditions. One warning sign is the breakdown of pedestrian flows
under congested conditions,
which causes stop-and-go waves and a further compression of the crowd \cite{PRE_makkah}. 
Another, very serious indicator of criticality
is the occurence of high values of the ``pressure'', which relates to the occurence of
turbulent crowd motion \cite{PRE_makkah}. This dangerous
dynamics of the crowd, which may cause people to fall, can be well seen in accelerated videos.

\subsection{Critical Discussion of the Obtained Results}

Some of our findings, including the extreme densities of up to 10 persons per square meter and more, the finite average speeds even at those densities, as well as the stop-and-go flows and turbulent crowd motion, may be hard to believe. Let us, therefore, in the following address some questions regarding the reliability of our study:
\begin{enumerate}
\item {\bf Reliability of video tracking:} The evaluation of dense crowds is certainly a difficult challenge, and the tracking of individual people is often not possible over long distances. However, this is also not needed to determine the densities, speeds and flows.\footnote{If a head is identified for the first time at a certain position, we say that pedestrian $i$ has a trackability 
level of $1$, and generally, if it holds for $n$ successive frames, the pedestrian 
has a trackability level of $n$. With this approach, an automatically identified head will be 
more certainly a real head, the higher the trackability level is. It is possible to require a
minimum trackability level $n_{\rm min}$ to match manual counts well. An average 
trackability level of at least $4$ turns out to be a good value for generating stable measurements.}
Our method is based on sophisticated image enhancement and pattern recognition methods, and it was independently evaluated, i.e. we did not have any influence on the manual counting procedure. When varying the time interval ${\cal T}$ (in minutes) between manual counts, the variance of the relative error ${\cal E}$ between automated and manual counts was given by 
\begin{equation}
{\cal E} =  D {\cal T}
\end{equation} 
with ${\cal D} = 0.0015 \pm 0.0003$ 
and a correlation coefficient ${\cal R}^2 = 0.72$. Therefore, the variance of the relative error grows with the sampling time interval ${\cal T}$ approximately according to a diffusion law (which represents a ``random walk'' of flow measurements over time). For ${\cal T}=10$ minutes, we have a standard deviation of $\sqrt{0.015} =12$ percent, which was considered to be acceptable by the evaluators. Note that this deviation does not only reflect errors in the automated counting procedure, but also variations in the measured quantities over the time interval between successive manual counts. Therefore, the error of the automated counts is much smaller,  and the proposed safety margin of 30\% should be large enough to take into account both, temporal variations in the flow and measurement errors. 
\par
Furthermore, note that the ``error bars'' shown in Fig. \ref{average} primarily reflect the spatial and temporal {\it variability} of the flow, and not the measurement error. For vehicle traffic, for example, it is a well-known fact that flow measurements at medium and high densities vary strongly in time \cite{ReviewofModernPhysics}. It is, therefore, likely that {\it all} original measurements of pedestrian flows have a large degree of variability, if the flow measurements for a given density are not averaged over. This variability is a matter of the time gap distribution \cite{ReviewofModernPhysics,TransSci}, which in turn is expected to depend on factors like walking individually or in groups, or the heterogeneity of the crowd with respect to culture and age.
\par
The accuracy of the global {\it density} measurements can be estimated from the distribution of {\it local} densities (see Fig. \ref{distribution}). When the density in an area of 20m$\times$\mbox{15m}  =  300m$^2$ is determined from local density measurements with $R=1$m and $A_R = 3.14$m$^2$, we have about 100 non-overlapping local density measurements. Averaging over them reduces the standard deviation by $\sqrt{100} = 10$ as compared to the one in Fig. \ref{distribution}, which results in a statistical error of less than 5\%  for the global density. Compared to the variation of the flow, this is negligible, so that it is justified to drop density-related error bars in Figs. \ref{average} and \ref{spatial_and_temporal_plots}.
\begin{figure}[!htbp]
\begin{center}
\includegraphics[width=8.5cm]{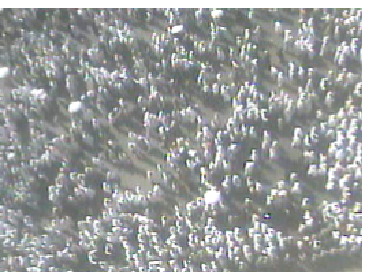}\\[2mm]
\includegraphics[width=8.5cm]{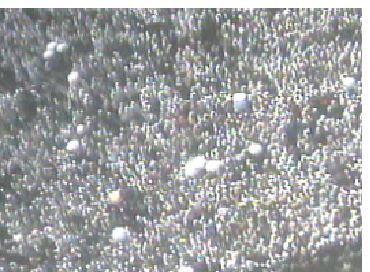}
\end{center}
\caption[]{Snapshots taken in the video-recorded area for different densities. It can be seen that the recordings were made from high altitude with little distortion to correct for. The pictures show that the distribution of people is not at all homogeneous. Instead, the local densities vary considerably. One can also see some umbrellas used for protection from the sun.} 
\label{example}
\end{figure}
\item {\bf Extreme densities:} Local densities of up to 10 persons per square meter or slightly more were not only determined from our video analysis. By evaluation of photographs and independently from us, they were obtained before. Moreover, considerations based on projected body areas suggest that densities upto 11 persons per square meters would be possible in principle  \cite{Still}, which is consistent with our results.
Furthermore, Mori's and Tsukaguchi's data \cite{Mori} come close to our measurements, when the scaling analysis of Fig. \ref{NEW} is applied. The importance of the body size distribution can also be seen in Refs.  \cite{Pheasant,Teknomo,Pauls}.
\item {\bf Flow-density relationship:} More important than the maximum possible densities (which should be avoided anyway) are the maximum global flow (as it determines the capacity of a pedestrian facility), and the density at which the flow starts to drop (which gives rise to congestion and a further increase of the density). It should be stressed that an occurence of the maximum flow should be avoided for two reasons: first, in order to prevent a breakdown of the flow and second, because the density at which the maximum flow is reached is not comfortable to pedestrians anymore \cite{Fruin2}. 
\par
Note that both, our local and global flow-density relations look different from the fundamental diagrams reported in the literature. In order to check the plausibility of our findings, we have scaled our data by the maximum density $\rho_{\rm max}$ (which depends on the projected body area) and by the average desired velocity $V_0$ (which is smaller for pilgrims than usual as they often walk in large groups including people aged 50 years and older). Our scaled speed-density data shown in Fig. \ref{NEW}a agree well with other measurements, which indicates their compatibility, while we find deviations for the flow-density  data at spatial occupancies of 0.6 and higher (see Fig. \ref{NEW}b). These deviations can have different origins: In some cases, fundamental diagrams have been determined by fitting speed-density data and multiplying them with the density. The resulting fit curve will not necessarily represent flow-density data well, as the flow-density relationship is sensitive to deviations in the speed-density relationship. Furthermore, the data for the Hajj are expected to differ from measurements for other situations in which dense crowds do not move (e.g. on the platform of a subway station or in front of a stage). Therefore, the question is whether the main distinguishing feature, the second increase in the flow at extreme densities, is an artifact of our evaluation method or represents a real fact.
\item {\bf Second increase in the flow at extreme densities:} This increase results from the measurement of finite average speeds even at extreme densities. In fact, it can be observed in the evaluated video recordings that (apart from intermediate stops reflecting stop-and-go motion) people kept moving at all observed densities, even after the accident occured and some areas were difficult to pass. People try to get out of areas of extreme density---this is why the average speed stays finite.
\par
It should be noted that the second increase of the flow has important implications for the dynamics \cite{Colombo}. In particular, it implies the coexistence of forward and backward moving shock waves. The minimum of the flow-density curve at finite densities and crowd turbulence are two closely related properties. And as crowd turbulence is well visible in the video recordings, when played in fast-forward mode (see the videos at http:/$\!$/www.trafficforum.org/crowdturbulence),
a second increase in the density is actually {\it expected} to exist.
\item {\bf Stop-and-go motion and crowd turbulence:} Both phenomena have not only been observed once and in a single location. Stop-and-go waves can occur in large spatial areas and persist over many minutes, and we have seen the phenomena in recordings of different locations. The same applies to crowd turbulence. It has not only occured in January 2006, but also in previous years, and the observation is consistent with what Fruin has reported \cite{Frui}: 
``At occupancies of about 7 persons per square meter the crowd becomes almost a fluid mass.
Shock waves can be propagated through the mass, sufficient to ... 
propel them distances of 3 meters or more. ... 
Access to those who fall is impossible.'' This visualizes the conditions in extremely dense crowds quite well, and it is compatible with the location of the minimum in our flow-density data.
\item {\bf Local vs. global dependencies:} It is clear that locally measured quantities vary more than global ones. If one averages over many local measurements, the maximum density and speed or flow are, of course, reduced. Nevertheless, according to Fig. \ref{NEW}b, the local and global curves agree surprisingly well, when the densities and speeds are scaled by the respective maximum values $V_0$ and $\rho_{\rm max}$. As the local and global quantities have been determined in different ways, this is a good reason to trust the corresponding data. 
\item {\bf Alarms of critical crowd conditions:}
The Hajj is in many ways an extreme event, with potentially critical situations occuring at different locations. If alarms are given whenever the global density exceeds a certain density defined to be critical, it is difficult to take specific actions. Such alarms may continue over many hours. The pressure-based method, in contrast, would give a warning of potentially turbulent crowd motion over a short time period, and would indicate critical locations. It is no problem if the critical pressure (here: 0.02/s$^2$)  is shortly crossed before crowd turbulence takes over in a large area. This gives security forces a bit more time to prepare counter actions. 
\par
Note, however, that counter actions should be taken much earlier, 
at latest when stop-and-go waves set in, as this indicates a breakdown of the flow. It should also be underlined that the critical flow of 0.8m/s \cite{PRE_makkah} and the critical pressure of 0.02/s$^2$ depend on the parameters of the measurement method, in particular the choice of $R$, and it may depend on other factors like the kind of crowd as well. As a consequence, these critical thresholds need to be properly calibrated. In fact, rather than an automated alarm system we propose to view accelerated video sequences, which allows one to easily discover areas where a smooth flow is disturbed, where stop-and-go waves appear or where even crowd turbulence occurs. Furthermore, it would be helpful to enhance surveillance videos by overlaying additional information like the intensity of the crowd pressure, as illustrated by the video at
http:/$\!$/www.trafficforum.org/crowdturbulence/pressure\_video.mpg 
\end{enumerate}

\subsection{Some Measures to Improve Crowd Safety}

In the interest of crowd safety, densities higher than 3-4 persons per square meter in large crowds and particularly the onset of stop-and-go waves or crowd turbulence must be avoided. Therefore, a combination of the following measures
is recommended:
\begin{enumerate}
\item {\it Design:} The infrastructure should be designed in a way that \underline{no} bottlenecks or objects (e.g. luggage) will obstruct the flow. This can be quite challenging, if the usage patterns are changing. In particular, it must be avoided that the outflow capacity is smaller than the inflow capacity. Accumulations of large crowds should be avoided. If this is impossible, the outflow capacities must be dimensioned such that the time to evacuate the area is much shorter than the time to fill it. (It should not exceed a few minutes.) Furthermore, a system of emergency routes or, more generally, a ``valve system'' is required to be able to reduce pressures in certain areas of the system, where needed. 
\item {\it Operation:} The infrastructure should be operated in a way that avoids counterflows or intersecting flows. Even merging flows may  cause serious problems. Flows of people should be re-balanced if there is a large utilization in certain parts of the system while there is still available capacity in others. Obstacles (including people blocking the ways) should be removed from areas intended for moving. Some ways should be reserved for emergency operation and protected from public access.
\item {\it Monitoring:} Areas of accumulation or any possible conflict points (including crossing flows or bottlenecks, if these are really unavoidable in the system) must be monitored during highly frequented time periods. To support the job of the monitoring crew and maneuver security forces to the right places, it is helpful to display additional information in the surveillance videos, visualizing, for example, the density, the flow, and/or the crowd pressure.
\item {\it Crowd Management:} In certain events, the flow must be suitably limited to the safe capacity of the system (which should consider a safety margin of 30\%). This may be done by applying a scheduling program, which is a plan regulating the timing and routing of groups of people. The compliance with the scheduling program must be carefully monitored (e.g. by control points and/or GPS tracking), and deviations from it must be counter-acted (e.g. by fines). Moreover, an adaptive re-scheduling should be possible in order to respond properly to the actual conditions in the system. It is even more favorable to have a simulation tool for the prediction of the flows in the system. This would allow an anticipative crowd management. 
\item {\it Contingency Plans:} For situations, where the system enters a critical state for whatever reason (e.g a fire, an accident, violent behavior, or bad weather conditions), one needs to have detailed contingency plans, which must be worked out and exercised in advance.
\end{enumerate} 
The above is a list of only {\it some} of the measures that can be taken to improve crowd safety. For further measures see Refs. 
\cite{Still,control1,control2,control3,control4,control5,control6,control7}.

\subsection*{Acknowledgments}

The authors are grateful for the partial financial support by 
the DFG grant He 2789/7-1 and the ETH Research Grant CH1-01 08-2.
They would like to thank the STESA staff for converting the large number of video recordings,
 and the MoMRA staff for doing the manual counting.
Furthermore, the authors appreciate fruitful discussions with Martin Treiber, who proposed the local density measure,
and with Karl Walkow, who collected and digitized empirical pedestrian data from other
sources for comparison with our measurements.


\begin{thebibliography}{99}
\bibitem{Schadschneider}
Burstedde, C., Klauck, K., Schadschneider, A., Zittartz, J.,
Simulation of pedestrian dynamics using a two-dimensional cellular automaton, 
{\em Physica A} {\bf 295} (4) (2001) 507--525.

\bibitem{Colombo}
Colombo, R. M. and Rosini, M. D. 
Pedestrian flows and non-classical shocks,
{\it Mathematical Methods in the Applied Sciences}   {\bf 28}(13), 1553-1567 (2005). 

\bibitem{ants}
Dussutour, A., Fourcassi\'{e}, V., Helbing, D., and Deneubourg, J.-L.,
Optimal traffic organization in ants under crowded conditions,
\textit{Nature} {\bf 428} (2004) 70--73.

\bibitem{SomeSuitableBookorReview}
Epstein, J., Axtell, R.,
{\em Growing Artificial Societies. Social Science from the Bottom Up} (Brookings Institution/MIT Press, Washington, DC, 1996).

\bibitem{Fruin2}
Fruin, J. J.,
Designing for pedestrians: A level-of-service concept,
\textit{Highway Research Record} {\bf 355} (1971) 1--15. 

\bibitem{Frui} J. J. Fruin, 
The causes and prevention of crowd disasters, 
in  \textit{Engineering for Crowd Safety}, edited by R. A. Smith and J. F. Dickie 
(Elsevier, Amsterdam, 1993), pp. 99-108.

\bibitem{control1}
Government of Canada, 
{\em Emergency preparedness guidelines for mass, crowd-intensive events},
see \verb|http://ww3.psepc-sppcc.gc.ca/research/resactivites/|
\verb|planPrep/en_crowd/1993-D011_e.pdf| (1994).
 
\bibitem{HakenSynergetics}
Haken, H.,
{\em Synergetics. an introduction} (Springer, Berlin, 1983).

\bibitem{control6}
Health and Safety Executive,
{\em Steps to Risk Assessment. Case Studies}
(HSE Books, Norwich, 1998).

\bibitem{control3}
Health and Safety Executive, 
{\em The Event Safety Guide}
(HSE Books, Norwich, 1999).
 
\bibitem{control4}
Health and Safety Executive, 
{\em Managing Crowds Safely}
(HSE Books, Norwich, 2000).

\bibitem{Sociodynamics}
Helbing, D.,
{\em Quantitative Sociodynamics. Stochastic Methods and Models of Social Interaction Processes}
(Kluwer Academic, Dordrecht, 1995).

\bibitem{ReviewofModernPhysics}
Helbing, D.,
Traffic and related self-driven many-particle systems. 
{\em Reviews of Modern Physics} {\bf 73} (2001) 1067--1141.

\bibitem{TransSci}
Helbing, D., Buzna, L., Johansson, A., and Werner, T.,
Self-organized pedestrian crowd dynamics: Experiments, simulations, and design solutions, 
\textit{Transportation Science} {\bf 39}(1) (2005) 1--24.

\bibitem{helbing2002}
Helbing, D., Farkas, I., Moln´ar, P., and Vicsek, T.,
Simulation of pedestrian crowds in normal and evacuation situations,
in {\em Pedestrian and Evacuation Dynamics}, M. Schreckenberg and S. D. Sharma (eds), 21--58 (Springer-Verlag, Heidelberg, 2002).

\bibitem{panic}
Helbing, D., Farkas, I., and Vicsek, T.,
Simulating dynamical features of escape panic,
{\it Nature} {\bf 407} (2000) 487--490.

\bibitem{PRE_makkah}
Helbing, D., Johansson, A., and Al-Abideen, H. Z.,
The dynamics of crowd disasters: An empirical study,
\textit{Physical Review E} {\bf 75}, 046109 (2007).

\bibitem{BottleneckPRL}
Helbing, D., Johansson, A., Mathiesen, J., Jensen, M.H., and Hansen, A.,
Analytical approach to continuous and intermittent bottleneck flows,
{\it Physical Review Letters} {\bf 97}, 168001 (2006).

\bibitem{HelbMoln1995}
Helbing, D. and Moln\'{a}r, P.,
Social force model for pedestrian dynamics,
{\it Physical Review E} {\bf 51} (1995) 4282--4286.

\bibitem{hoogendoorn2005} 
Hoogendoorn, S. P. and Daamen, W.,
Pedestrian behavior at bottlenecks,
{\it Transportation Science} {\bf 39}(2) (2005) 147--159.

\bibitem{HoogendoornEmpirical}
Hoogendoorn, S. P., Daamen, W., and Bovy, P.H.L.,
Extracting microscopic pedestrian characteristics from video data,
{\it 82nd Transportation Research Board Annual Meeting} (CD-ROM, Washington DC: National Academy
Press, 2003).

\bibitem{control7}
Independent Street Arts Network, 
{\em Safety Guidance for Street Arts, Carnival, Processions, and Large-Scale Performances}
(Independent Street Arts Network, London, 2004).

\bibitem{EvacuationWithNagatani}
Isobe, M., Helbing, D., and Nagatani, T.,
Experiment, theory, and simulation of the evacuation of a room without visibility. 
{\em Physical Review E} {\bf 69} (2004) 066132.

\bibitem{johansson_phd}
Johansson, A.,
Data-driven Modeling of Pedestrian Crowds, Ph.D. thesis,
in preparation,
Dresden University of Technology, Dresden, (2008).

\bibitem{Kerridge}
Kerridge, J., Keller, S., Chamberlain, T., and Sumpter, N.,
Collecting Pedestrian Trajectory Data In Real-time,
in {\it Pedestrian and Evacuation Dynamics 2005},
N. Waldau {\it et al.} (eds),
27--39 (Springer-Verlag, Heidelberg, 2007).

\bibitem{Kretz}
Kretz, T., Gr\"unebohm, A., and Schreckenberg, M.,
Experimental study of pedestrian flow through a bottleneck,
{\textit J. Stat. Mech} P10014 (2006).

\bibitem{Mori}
Mori, M. and Tsukaguchi, H.,
A new method for evaluation of level of service in pedestrian facilities, 
{\it Transportation Research A} {\bf 21}(3) (1987) 223--234 .

\bibitem{MehdisPNAS}
Moussaid, M., Garnier, S., Helbing, D., Johansson, A., and Theraulaz, G.,
Experimental measurements of human interactions in space and time,
Submitted.

\bibitem{control2}
Musterversammlungsst\"attenverordnung 2005,
see\\ \verb|http://www.versammlungsstaettenverordnung.de/vstaettv_neu/|
\verb|bundeslaender/downloads/MusterVstaettV2005/MVStaettV_2005.pdf|
(2005).

\bibitem{Pauls}
Pauls, J. L., Fruin, J. J., and Zupan, J. M.,
Minimum stair-width for evacuation, overtaking movement and counterflow -- technical bases and suggestions for the past, present and future,
in {\it Pedestrian and Evacuation Dynamics 2005},
N. Waldau {\it et al.} (eds),
57--69 (Springer-Verlag, Heidelberg, 2007).

\bibitem{Pheasant}
Pheasant, S. T.,
{\em Bodyspace: Anthropometry, Ergonomics, and the Design of Work}
(Taylor and Francis, London, 1998).

\bibitem{Polus}
Polus, A., Schofer, J. L., Ushpiz, A.,
Pedestrian flow and level of service,
{\it Journal of Transportation Engineering} {\bf 109} (1983) 46--56.

\bibitem{Predtechenskii}
Predtechenskii, V. M. and Milinskii, A. I.,
{\em Planning for Foot Traffic Flow in Buildings}
(Amerind, New Delhi, (1978).

\bibitem{mice} 
Saloma, C., Perez, G. J., Tapang, G., Lim, M., and Palmes-Saloma, C.,
Self-organized queuing and scale-free behavior in real escape panic,
\textit{PNAS} \textbf{100}(21) (2003) 11947--11952.

\bibitem{schadschneider_encyclopedia}
Schadschneider, A., Klingsch, W., Kluepfel, H., Kretz, T., Rogsch, C., and Seyfried, A.
Evacuation dynamics: empirical results, modeling and applications,
arXiv:0802.1620 (2008).

\bibitem{BrownianAgents}
Schweitzer, F.,
{\em Brownian Agents and Active Particles. Collective Dynamics in the Natural and Social Sciences} (Springer, Berlin, 2003).

\bibitem{Seyfried}
Seyfried, A., Steffen, B., Klingsch, W., Boltes, M.,
The fundamental diagram of pedestrian movement revisited,
\textit{J. Stat. Mech.} P10002 (2005).

\bibitem{Still}
Still, K.,
Crowd Dynamics, Ph.D. thesis,
University of Warwick, Warwick, (2000).

\bibitem{Teknomo} 
Teknomo, K.,
Microscopic Pedestrian Flow Characteristics: Development of an Image Processing Data Collection and Simulation Model, 
Ph.D. thesis, Tohoku University Japan, Sendai, (2002).

\bibitem{control5}
The Scottish Office, 
{\em Guide to Safety at Sports Grounds}
(The Stationery Office, Norwich, 4th edition, 1997).

\bibitem{Weidmann}
Weidmann, U.,
{\it Transporttechnik der Fu{\ss}g\"anger} (Schriftenreihe des Institut f\"ur Verkehrsplanung, Transporttechnik, Stra{\ss}en- und Eisenbahnbau {\bf 90}, ETH Z\"urich, Z\"urich, 1993).

\bibitem{RichardWiseman}
Wiseman, R.
{\em Pace of Life},\\
\verb|http://www.paceoflife.co.uk/| (2008).
 
\bibitem{turbulenceModel}
Yu, W. and Johansson, A.,
Modeling crowd turbulence by many-particle simulations,
{\it Phys. Rev. E} {\bf 76}, 046105 (2007) 

\bibitem{rimea}
Richtlinie f\"ur Mikroskopische Entfluchtungsanalysen (in German)
The RiMEA Project\\
\verb|http://www.rimea.de/downloads.html| (2008)

\bibitem{supplement}
{\em Supplementary web page},
\verb|http://www.trafficforum.org/crowdturbulence| (2007)

\end{thebibliography}
\end{document}